%% file: main.tex
\documentclass[10pt,conference]{IEEEtran}
\usepackage{cite}
\usepackage{amsmath,amssymb,amsfonts}
\usepackage{algorithmicx}
\usepackage{algcompatible}
\usepackage{graphicx}
\usepackage{textcomp}
\usepackage{xcolor}
\usepackage[hyphens]{url}
\usepackage{fancyhdr}
\usepackage{hyperref}

\newcommand{\hpcayear}{2026}

\newcommand{\hpcasubmissionnumber}{171}
\title{AQPIM: Breaking the PIM Capacity Wall for LLMs with In-Memory Activation Quantization}

\def\hpcacameraready{} 

\newcommand\hpcaauthors{Kosuke Matsushima$^\dagger$, Yasuyuki Okoshi$^\dagger$, Masato Motomura$^\dagger$, and Daichi Fujiki$^\dagger$}
\newcommand\hpcaaffiliation{Institute of Science Tokyo$^\dagger$}
\newcommand\hpcaemail{\{matsushima.kosuke, okoshi.yasuyuki, motomura, dfujiki\}@artic.iir.isct.ac.jp}

\usepackage{cleveref}
\usepackage{caption}
\usepackage{algorithm}
\usepackage{algpseudocode}
\usepackage{tablefootnote}
\usepackage{multirow}
\usepackage{graphicx}
\usepackage{subcaption}
\usepackage{xspace}
\usepackage{tabularx}
\usepackage{array}
\usepackage{graphicx}
\usepackage{makecell}
\usepackage{todonotes}
\usepackage{pifont}
\usepackage{comment}
\usepackage{hyperref}

\usepackage[T1]{fontenc}
\usepackage[utf8]{inputenc}

\crefname{figure}{Figure}{Figures}
\crefname{table}{Table}{Tables}
\crefname{equation}{Eq.}{Eqs.}
\newcommand{\bhline}[1]{\noalign{\hrule height #1}}

\newcolumntype{C}{>{\centering\arraybackslash}X}
\newcolumntype{P}[1]{>{\centering\arraybackslash}p{#1}}

\newcommand{\AQPIM}{AQPIM\xspace}

\newcommand{\RB}[1]{\todo[color=orange!40]{#1}}
\newcommand{\RBtext}[1]{{\color{orange}#1}}
\renewcommand{\RB}[1]{}
\renewcommand{\RBtext}[1]{#1}

\algnewcommand{\LineComment}[1]{\Statex \hskip\algorithmicindent \(\triangleright\) #1}

\begin{document}

\newpage

\input{hpca-template}

\setcounter{page}{1}


\input{sections/0_abstract}
\input{sections/1_introduction}
\input{sections/2_background}

\input{sections/3_proposal}
\input{sections/4_evaluation}
\input{sections/6_conclusion}

\section*{Acknowledgements}
This work was supported by JSPS KAKENHI Grant Numbers JP25K03092, JP23H05489, JST BOOST Grant Number JPMJBY24G7, JST PRESTO Grant Number JPMJPR22P7, and JST ALCA-Next Grant Number JPMJAN24F3.


\bibliographystyle{IEEEtranS}
\bibliography{refs}

\end{document}

%% file: hpca-template.tex

\author{
  \ifdefined\hpcacameraready
    \IEEEauthorblockN{\hpcaauthors{}}
      \IEEEauthorblockA{
        \hpcaaffiliation{} \\
        \hpcaemail{}
      }
  \else
    \IEEEauthorblockN{\normalsize{HPCA \hpcayear{} Submission
      \textbf{\#\hpcasubmissionnumber{}}} \\
      \IEEEauthorblockA{
        Confidential Draft \\
        Do NOT Distribute!!
      }
    }
  \fi 
}

\fancypagestyle{camerareadyfirstpage}{%
  \fancyhead{}
  \renewcommand{\headrulewidth}{0pt}
  \fancyhead[C]{
    \ifdefined\aeopen
    \else
      \ifdefined\aereviewed
      \else
      \ifdefined\aereproduced
      \else
    \fi 
    \fi 
    \fi 
    \ifdefined\aeopen 
      \includegraphics[width=12mm,height=12mm]{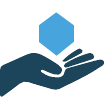}
    \fi 
    \ifdefined\aereviewed
      \includegraphics[width=12mm,height=12mm]{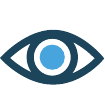}
    \fi 
    \ifdefined\aereproduced
      \includegraphics[width=12mm,height=12mm]{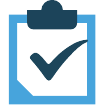}
    \fi
  }
  \fancyfoot[C]{}
}
\fancyhead{}
\renewcommand{\headrulewidth}{0pt}

\maketitle

\ifdefined\hpcacameraready 
  \thispagestyle{camerareadyfirstpage}
  \pagestyle{empty}
\else
  \thispagestyle{plain}
  \pagestyle{plain}
\fi

\newcommand{\hpcaheight}{0mm}
\ifdefined\eaopen
\renewcommand{\hpcaheight}{12mm}
\fi

%% file: sections/0_abstract.tex
\begin{abstract}
Processing-in-Memory (PIM) architectures offer a promising solution to the memory bottlenecks in data-intensive machine learning, yet often overlook the growing challenge of activation memory footprint. Conventional PIM approaches struggle with massive KV cache sizes generated in long-context scenarios by Transformer-based models, frequently exceeding PIM's limited memory capacity, while techniques like sparse attention can conflict with PIM's need for data locality. Existing PIM approaches and quantization methods are often insufficient or poorly suited for leveraging the unique characteristics of activations. This work identifies an opportunity for PIM-specialized activation quantization to enhance bandwidth and compute efficiency.

We explore clustering-based vector quantization approaches, which align well with activation characteristics and PIM's internal bandwidth capabilities. Building on this, we introduce AQPIM, a novel PIM-aware activation quantization framework based on Product Quantization (PQ), optimizing it for modern Large Language Models (LLMs).  By performing quantization directly within memory, AQPIM leverages PIM's high internal bandwidth and enables direct computation on compressed data, significantly reducing both memory footprint and computational overhead for attention computation. AQPIM addresses PQ's accuracy challenges by introducing several algorithmic optimizations. Evaluations demonstrate that AQPIM achieves significant performance improvements, \RBtext{drastically reducing\RB{C3} of GPU-CPU communication that can account for 90$\sim$98.5\% of decoding latency, together with 3.4$\times$ speedup over a SOTA PIM approach.
}

\end{abstract}

%% file: sections/1_introduction.tex
\section{Introduction}
\label{sec:Introduction}

Large-scale machine learning models, such as Large Language Models (LLMs), have demonstrated remarkable capabilities~\cite{hadi2023large, geminiteam2024gemini15unlockingmultimodal, grattafiori2024llama3herdmodels, Jiang2023mistral7b, guo2025deepseekr1, liu2024deepseek}. However, their escalating data demands are driving up the cost of data movement, which challenges operational efficiency and sustainability~\cite{iea_energy_2025}. While conventional optimizations have focused on static parameters such as weights~\cite{xiao2024efficient, spa_child2019generating, spa_beltagy2020longformer, spa_zaheer2020big, sze2020efficient, liu2022s2ta, chen2019eyeriss, parashar2017scnn}, the primary challenge has shifted to the dynamically generated activation (KV) cache, whose memory footprint grows with the volume of contextual data during inference. This trend is further accelerated by the demand for complex reasoning tasks that require longer contexts~\cite{guo2025deepseekr1}. Unlike pre-defined weights, these dynamic activations demand real-time optimizations and close collaboration between hardware and software to be managed efficiently~\cite{sheng2023flexgen, zhao2024alisa}.

The emergence of Processing-in-Memory (PIM) has opened up significant avenues for mitigating the data movement problem~\cite{fujiki2021near, kim2021aquabolt, lee2021hardware, fujiki2023mvc, park2024attacc!, heo2024neupims, eckert2018neural, fujiki2019duality}. By executing computations closer to or within memory, PIM enables a spectrum of optimizations, from accelerating localized arithmetic kernels near banks~\cite{kim2021aquabolt, lee2021hardware, park2024attacc!, heo2024neupims} to performing complex matrix operations directly in memory~\cite{eckert2018neural, fujiki2019duality, shafiee2016isaac, chi2016prime}. 
Recently, PIM architectures have been increasingly applied to optimize memory-bound attention mechanisms in Transformers and LLMs~\cite{park2024attacc!, heo2024neupims}. A critical application is the autoregressive decoding phase of LLMs, where the GEMV operation between a new token and the KV cache is memory-bound and performing it in-situ effectively alleviates the performance bottleneck~\cite{park2024attacc!, heo2024neupims}. 

Despite the potential of PIM, several challenges and missed opportunities remain unaddressed. 
First, current PIM approaches primarily focus on computational gains by leveraging internal memory bandwidth for attention and matrix operations~\cite{park2024attacc!, heo2024neupims}, overlooking the critical challenge of memory capacity. Existing PIM architectures struggle with the massive KV cache sizes generated, especially in long-context scenarios (potentially hundreds of GBs)~\cite{Bai2024longbench, bulatov2024scalingtransformer1mtokens, sheng2023flexgen}. This frequently exceeds PIM's memory capacity, which is already reduced due to the huge density costs of implementing bank-level PIM~\cite{lee2021hardware}, creating a fundamental \emph{capacity wall}. Consequently, even SOTA accelerators require a prohibitively large number of devices to accommodate the entire context in PIM (e.g., 40 HBMs for short contexts~\cite{park2024attacc!}), exposing a critical scaling issue that current PIM designs overlook.

Conventional methods to mitigate this memory pressure are unfortunately ill-suited for the PIM paradigm. While techniques such as offloading or sparse attention can reduce memory footprint on conventional systems~\cite{sheng2023flexgen, zhao2024alisa, zhang2024pqcacheproductquantizationbasedkvcache, liu2024clusterkvmanipulatingllmkv, off_liu2024retrievalattention, off_wu2022memorizing}, these techniques are not effective for PIM due to their scattered access patterns~\cite{yuan2025nativesparse}, directly contradicting the need for data locality and contiguous memory access required for efficient computation in PIM~\cite{lee2021hardware, kim2021aquabolt, park2024attacc!}. 

This naturally leads to considering on-chip compression via quantization. However, this path reveals a critical roadblock for PIM. Implementing mainstream quantization schemes~\cite{liu2023KIVI, duanmu2024skvq, hooper2024kvquant, he2024zipcache, xiao2023smoothquant, lin2024duquant, wenqi2024omniquant, quant_zhao2024atom, quant_lin2024qserve} directly in PIM requires additional hardware to existing FP16 MAC units for numerical scaling and controls, which incurs a prohibitive area overhead ($\sim$126\% just for FP16+INT32~\cite{hbmpim}), destroying the memory density. Thus, simply adding quantization logic to PIM would not be a viable solution.
Moreover, existing non-PIM hardware-based quantization methods~\cite{gholami2022survey, zhu2024survey} have primarily focused on value-by-value quantization, assuming the orthogonality of weights, but have failed to exploit the inherent locality and similarity in the context-dependent distributions of activations, missing significant compression opportunities.
Therefore, a significant research gap exists in optimizing KV quantization specifically for the distinct attributes of PIM architectures, i.e., high internal bandwidth, low off-chip bandwidth, and limited computational resources and area.



This paper presents \AQPIM, a novel PIM-aware activation compression technique that resolves these challenges through a synergistic algorithm-hardware co-design. Our approach is built upon Product Quantization (PQ)~\cite{Jégou2011productquantization}, a clustering-based method that effectively captures the inherent similarity and locality within activations. While clustering-based quantization has long been recognized for its efficiency, its significant bandwidth requirements have limited its applicability, making it impractical for online use in conventional architectures~\cite{yuan2025nativesparse}. The core insight of our work is to turn this long-standing challenge into a key opportunity: we repurpose PIM's massive and often underutilized internal bandwidth to service the intense demands of online clustering. This strategic move makes a superior but previously inaccessible algorithm practical, breaking the capacity wall without costly hardware modifications.

AQPIM's efficiency is realized through several key innovations. First, it transforms the expensive GEMV operation in attention into a sequence of efficient lookups and summations that operate directly on the compressed data, eliminating the need for dequantization and allowing the use of existing simple FP16 MAC units. The primary bottleneck of this lookup-based method—the random access penalty for retrieving centroid data—is solved by a HW co-design. We algorithmically ensure lookup tables reside within a single DRAM row and architecturally add minimal hardware for fast indirect addressing, turning random logical accesses into predictable row-buffer hits. Combined with further algorithmic improvements such as importance-aware clustering, AQPIM achieves high accuracy and performance.

The key contributions of this work are as follows:
\begin{itemize}
    \vspace{-1mm}
    \item A novel co-design that enables practical, high-fidelity online clustering-based quantization by leveraging PIM's massive internal bandwidth.

    \item A PIM-aware attention mechanism that computes directly on compressed data, resolving the critical random-access bottleneck through a tight co-design of page-aware clustering and minimal hardware support.
    
    \item Algorithmic enhancements and optimizations to maximize clustering accuracy under high compression ratios with minimal overhead.
    
    \item The fully integrated \AQPIM framework, \RBtext{demonstrating drastic reduction\RB{C3}\label{RB:C3} of GPU-CPU communication that can account for 90$\sim$98.5\% of decoding latency, together with 3.4$\times$ speedup over a SOTA PIM approach due to aggressively reduced KV cache size.} 
\end{itemize}

%% file: sections/2_background.tex
\section{Background and Motivation}
\label{sec:background}

\subsection{PIM Acceleration for LLM Inference}
\label{ssec:pim_acceleration_for_llm_inference}

The escalating demand for processing longer texts with LLMs has led to a significant expansion of their context window sizes.
Modern models like Llama 3~\cite{grattafiori2024llama3herdmodels}, GPT-4o~\cite{openai2025gpt4o} and Gemini 1.5~\cite{geminiteam2024gemini15unlockingmultimodal} support context windows of up to 128K, 128K, and 1M tokens, respectively. This trend substantially increases the memory access demands. Furthermore, recent advancements in LLMs enable them to generate longer outputs for complex tasks such as reasoning~\cite{reasoning_chen2022program, reasoning_wei2022chain, reasoning_gao2023pal, reasoning_gou2024tora}, where the model produces detailed explanatory tokens to clarify the logical process, leading to longer decoding output.

Processing long-context in LLM inference presents a critical bottleneck, particularly within the attention mechanism. Attention layers are frequently memory-bound, especially during the decoding phase. This is because each new token generation requires accessing the KV cache, which stores all previously generated tokens. As the context window grows, the KV cache size scales linearly, demanding significant memory bandwidth to fetch these large data structures for GEMV operations.

Recognizing this, prior PIM accelerators such as AttAcc!~\cite{park2024attacc!} and NeuPIMs~\cite{heo2024neupims} specifically target attention computation as the primary performance bottleneck for LLM inference, and apply PIM for memory-bound operations. 
AttAcc! proposes a heterogeneous system that combines the powerful computational capabilities of GPU with the high intra-memory bandwidth of PIM.
NeuPIMs, on the other hand, integrates NPU with PIM, aiming to maximize concurrent processing by employing a dual row buffer architecture for simultaneous data access and a sub-batch interleaving for fine-grained pipelining. 



However, existing PIM architectures struggle with a critical challenge: the sheer volume of KV caches generated, particularly in long-context scenarios, which can reach hundreds of GBs. This often surpasses PIM's memory capacity, especially since \textit{implementing bank-level PIM incurs costs that reduce memory density}. Without mitigation, this can lead to out-of-memory (OOM) crashes, making KV cache compression or offloading crucial for ensuring efficient and scalable performance.


\subsection{KV Cache Mitigation}

Several approaches have been proposed to reduce the memory footprint of the KV cache in the device memory: \emph{eviction}, \emph{offloading}, and \emph{quantization}. 

\textbf{Eviction}: 
Following both static
~\cite{xiao2024efficient, spa_child2019generating, spa_beltagy2020longformer, spa_zaheer2020big}
and dynamic eviction strategies
~\cite{zhang2023ho, li2024snapkv, spa_adnan2024keyformer, spa_oren2024transformers, spa_jiang2024minference},
eviction-based methods keep important tokens to enable lightweight attention computations.
StreamingLLM~\cite{xiao2024efficient} introduces a static token eviction rule that retains certain initial tokens, referred to as \textit{sink tokens}, along with a sliding window, since these tokens tend to produce high attention scores. 
SnapKV~\cite{li2024snapkv}
dynamically calculates token importance scores by aggregating recent attention scores during prefilling, and then selects the top-$k$ tokens based on these scores, where $k$ is the number of tokens to be retained.
As a result, KV cache becomes sparse, reducing both memory usage and computational cost in the attention mechanism. However, eviction-based approaches inherently suffer from irreversible token loss, leading to accuracy degradation, particularly in long-context scenarios.

\textbf{Offloading and Sparsification}: 
To mitigate this irreversible token loss, offloading methods 
~\cite{sheng2023flexgen, zhao2024alisa, zhang2024pqcacheproductquantizationbasedkvcache, liu2024clusterkvmanipulatingllmkv, off_liu2024retrievalattention, off_wu2022memorizing} 
preserve the entire KV cache within the memory hierarchy, including the host's main memory and storage. To minimize the overhead of accessing low-bandwidth memory, offloading methods often employ \emph{sparse attention}, which selects only the important tokens for attention computation, thereby reducing data transfer from low-bandwidth memory.  
Various approaches to vector similarity search, including 
PQ~\cite{zhang2024pqcacheproductquantizationbasedkvcache} and graph-based ANNS~\cite{off_liu2024retrievalattention}, are employed to identify important tokens, and by retrieving only a small subset of KVs from the host memory, they minimize data transfer overhead. However, these approaches suffer from the bandwidth limitation of the external memory and scattered memory access patterns. For example, NVIDIA's H100 GPU~\cite{nvidia2024h100} has approximately 26$\times$ lower bandwidth for CPU memory access compared to that of HBM.
In addition, sparse attention methods often struggle to maintain memory efficiency in Grouped-Query Attention (GQA) and Multiple-Query Attention (MQA) because the shared KV-cache requires accessing the union of KV selections from all query heads within a group, requiring high memory access~\cite{yuan2025nativesparse}.

\textbf{Quantization}: Quantization~\cite{gholami2022survey, zhu2024survey} has been widely explored for model compression in neural networks, and has also been applied to KV cache compression~\cite{liu2023KIVI, duanmu2024skvq, hooper2024kvquant, he2024zipcache, xiao2023smoothquant, lin2024duquant, wenqi2024omniquant, quant_zhao2024atom, quant_lin2024qserve}.
It can be categorized into uniform and non-uniform quantization depending on how the original value is mapped into a quantized value. 

Uniform quantization maps a real value into an integer by rounding to the nearest integer within fixed intervals.
A group of integer values is multiplied by a scaling factor to align the distribution of quantized values with the original.
Despite its small bandwidth requirements, it suffers from accuracy degradation, especially when applied to KV cache due to its outliers.
Existing works mitigate this problem by increasing the granularity of quantization~\cite{he2024zipcache, quant_lin2024qserve}, optimizing granularity depending on the target~\cite{liu2023KIVI}, or smoothing outliers~\cite{xiao2023smoothquant,wenqi2024omniquant,lin2024duquant}).
In addition, achieving inference acceleration often necessitates the quantization of model weights as well, which may result in accuracy degradation.

Non-uniform quantization~\cite{zadeh2020gobo, cluster_based_monkey, hooper2024kvquant, hu2025m} maps the original distribution into non-uniform datatypes.
KVQuant~\cite{hooper2024kvquant} determines the datatypes by using calibration data. It minimizes the quantization error in calibration datasets to obtain optimal quantization values.
M-ANT~\cite{hu2025m} introduces an adaptive numerical type that can support diverse distributions. This also leverages calibration datasets to determine the distribution of KV cache to avoid computational overhead.
However, the requirement of calibration datasets potentially leads to sub-optimal performance, especially when processing inputs with different distributions.

The cluster-based approach can mitigate this problem by adjusting quantization values to match the original distribution on the fly.
Rather than relying on calibration data, it directly computes quantization centroids from the input data, resulting in accuracy-optimal methods for a specified bit-width.
Despite its compression efficiency, it has been considered to be impractical due to the required bandwidth coming from the iterative process for centroid calculation~\cite{yuan2025nativesparse}.
Thus, its application is typically limited to weight-only quantization with a limited granularity of quantization~\cite{cluster_based_monkey, zadeh2020gobo}.

\begin{figure}[tb]
    \centering
    \includegraphics[width=.9\linewidth]{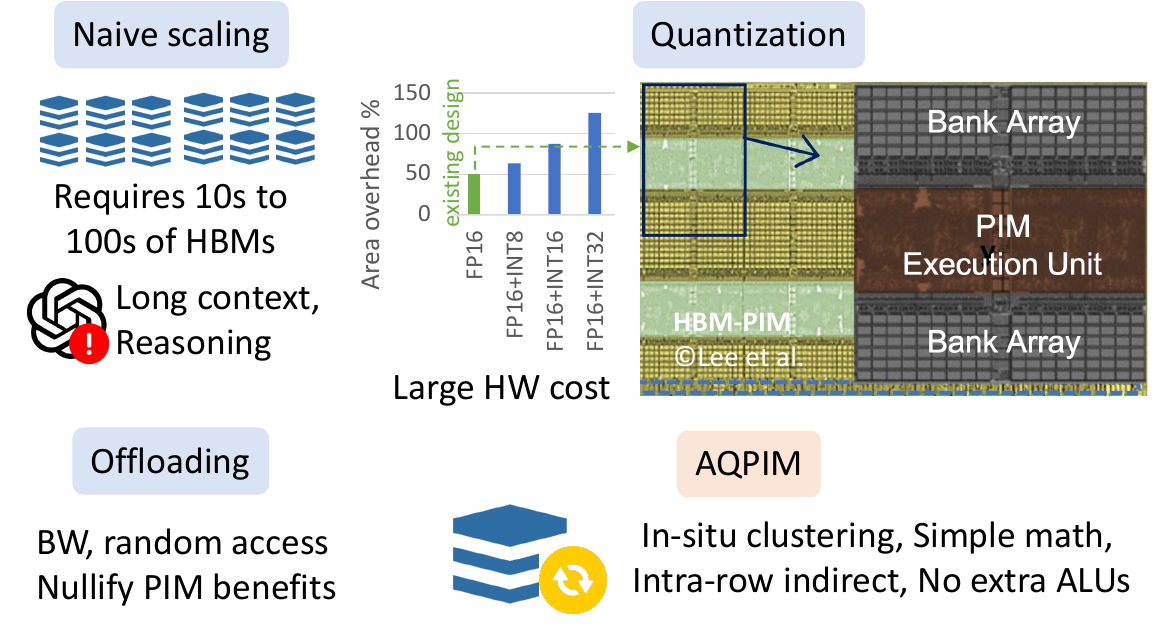}
    \caption{Scaling challenges of existing PIM designs for LLMs. The die photo is taken from the HBM-PIM paper~\cite{lee2021hardware}.}
    \vspace{-2mm}
    \label{fig:aqpim-concept}
    \vspace{-2mm}
\end{figure}

\subsection{Motivation}
\label{ssec:motivation}

\subsubsection{The PIM Capacity Wall and Quantization Dilemma}
While PIM architectures show promise for accelerating LLM inference, they face a fundamental \emph{capacity scaling problem}, as illustrated in~\Cref{fig:aqpim-concept}. Current PIM accelerators are primarily performance-focused, often assuming the entire KV cache fits within PIM's limited on-chip memory. This assumption quickly breaks down in long-context scenarios, where the KV cache can swell to hundreds of GBs. For instance, a SOTA accelerator like AttAcc!~\cite{park2024attacc!} already requires as many as 40 HBM-PIM devices to support even short contexts. This number becomes prohibitively large and economically unviable for the long-context scenarios targeted by modern LLMs.

A seemingly obvious solution---offloading the excess KV cache to the host memory hierarchy---is not a feasible remedy. The high communication overhead of traversing the PCIe bus would nullify PIM's performance gains. Furthermore, sparse attention techniques often used in conjunction with offloading introduce scattered, irregular memory access patterns that directly conflict with PIM's architectural need for data locality to achieve high utilization of its PEs.

This leads to the consideration of data compression techniques, such as quantization. However, implementing conventional quantization methods directly within PIM presents a critical dilemma. Mainstream quantization/dequantization schemes often rely on integer or mixed precision arithmetic (e.g., INT16/INT32 with FP16~\cite{quant_lin2024qserve}) and complex scaling operations. Integrating the necessary compute units for these operations into the already area-constrained bank-level PEs would incur a prohibitive area overhead. As noted in prior work~\cite{lee2021hardware, kim2021aquabolt}, simply adding these integer MAC units could increase the logic area from 50\% (FP16 only) to as much as 126\% (FP16+INT32), severely compromising memory density. 
Thus, simply adding new hardware to PIM is not a practical path forward. 



\subsubsection{Our Goals}
\begin{figure}[tb]
  \centering
  \includegraphics[width=\linewidth]{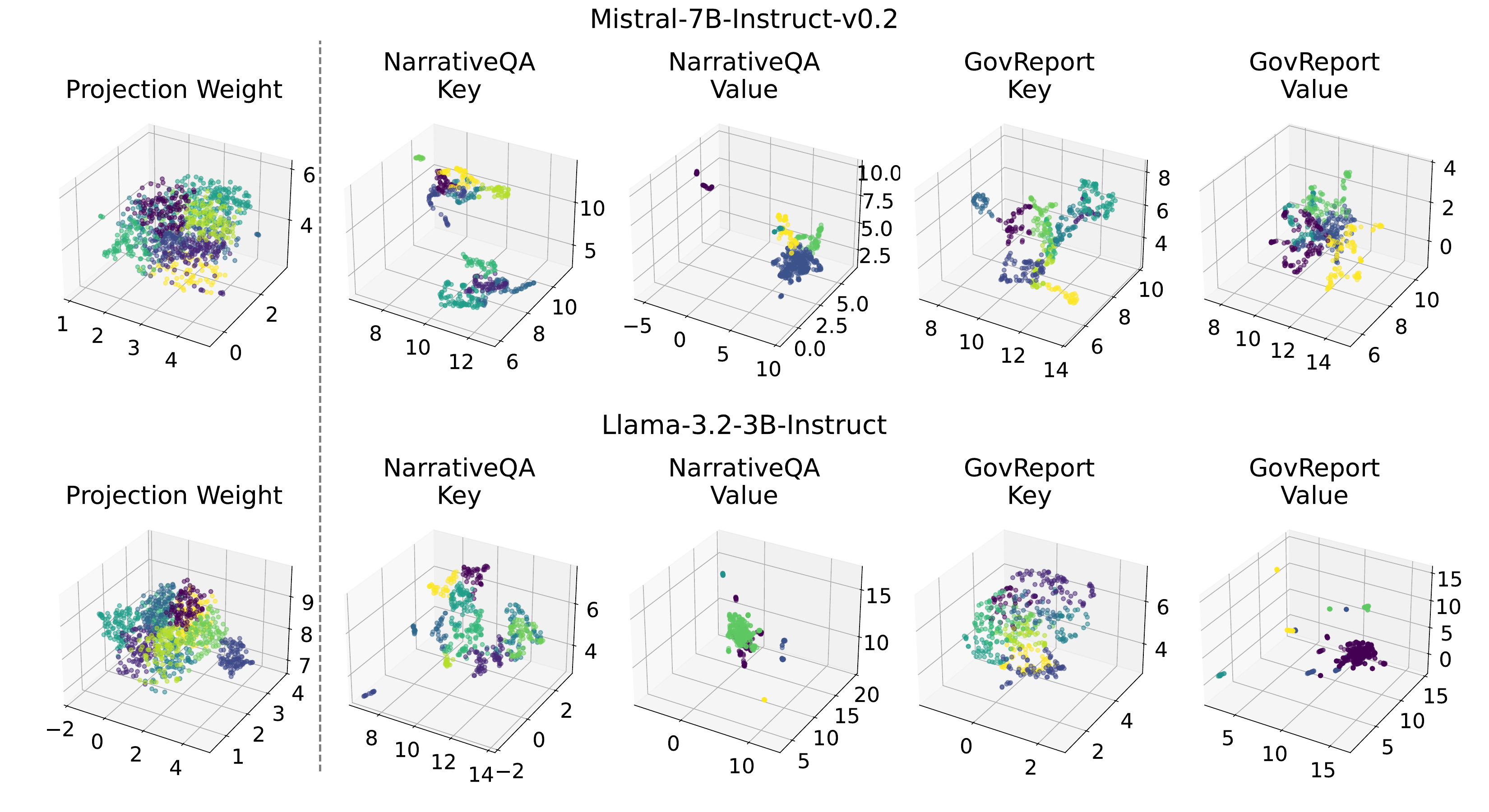}
  \caption{
  Locality within the projection weights (left-most) and
KV cache (right-four) visualized by UMAP~\cite{mcinnes2020umapuniformmanifoldapproximation}, using Mistral-7B-Instruct-v0.2~\cite{Jiang2023mistral7b} and Llama-3.2-3B-Instruct~\cite{grattafiori2024llama3herdmodels} running NarrativeQA~\cite{kociskyetal2018narrativeqa} and GovReport~\cite{huang-etal-2021-efficient}. 
\RBtext{We present the full results in~\cite{zenodo}.}
  }
  \label{fig:kv_locality}
  \vspace{-3mm}
\end{figure}

The challenges identified above define a clear set of principles that must guide the design of a truly practical and scalable PIM-based LLM accelerator. This work introduces AQPIM, a framework built upon the following core design goals:

\textbf{High-Ratio Compression with High Fidelity in PIM.} 
The primary goal is to break the capacity wall. This requires a compression technique that can drastically reduce the KV cache footprint while preserving model accuracy. Our key insight comes from observing the fundamental properties of activations. The distribution of activations is highly context-dependent, exhibiting significant \emph{locality and similarity}. This is visually demonstrated in \cref{fig:kv_locality} using UMAP~\cite{mcinnes2020umapuniformmanifoldapproximation}, a dimensionality reduction technique that preserves data's topological structure. As illustrated, key and value vectors exhibit a non-uniform distribution with tight clusters, unlike the more evenly distributed weight vectors. This inherent locality makes the KV cache particularly well-suited for clustering-based quantization such as Product Quantization (PQ), as clusters can naturally adapt to the underlying data distribution. 

\textbf{Synergistic Algorithm-Hardware Co-design.} 
The solution must actively leverage PIM's unique strengths, not just work around its constraints. While powerful, clustering-based quantization has been considered impractical for on-the-fly use due to its massive bandwidth demand in conventional systems. Our design turns this challenge into an opportunity by repurposing PIM's large, underutilized internal bandwidth to service the demands of online clustering. This synergy makes a superior but previously inaccessible algorithm practical, achieving both high compression and high accuracy.

\textbf{Efficient Computation with Minimal Area Overhead.} With online clustering made feasible, the next challenge is performing attention computation without introducing significant logic to the PIM PEs. We introduce a technique to transform the expensive GEMV operations into a sequence of localized lookups and summations of centroids, which is inherently suited for the simple FP16 MAC units already present in PIM. A key advantage of our approach is that it operates directly on the compressed data, eliminating the need for a separate dequantization. 
The two potential bottlenecks of this approach, i.e., random lookup latency and data growth, are solved via a tight algorithmic and architectural co-design. To eliminate lookup latency, we employ a \emph{page-aware windowed clustering algorithm}. This method maps tokens within a sliding context window to a compact set of centroids that are guaranteed to reside within a single DRAM row. Architecturally, we then introduce a minimal hardware enhancement (indirect addressing in the row buffer) to capitalize on this data locality, making every lookup a fast row-buffer hit. This, combined with an efficient page-aware strategy to update token indices as the context grows, makes the entire computation efficient on existing hardware.

AQPIM is the realization of these design goals, offering a comprehensive framework that enables efficient and flexible activation quantization for next-generation LLM inference on PIM architectures.

%% file: sections/3_proposal.tex
\section{\AQPIM}

\begin{figure*}[tb]
    \centering
    \begin{subfigure}{0.66\linewidth}
        \centering
        \includegraphics[height=4.3cm]{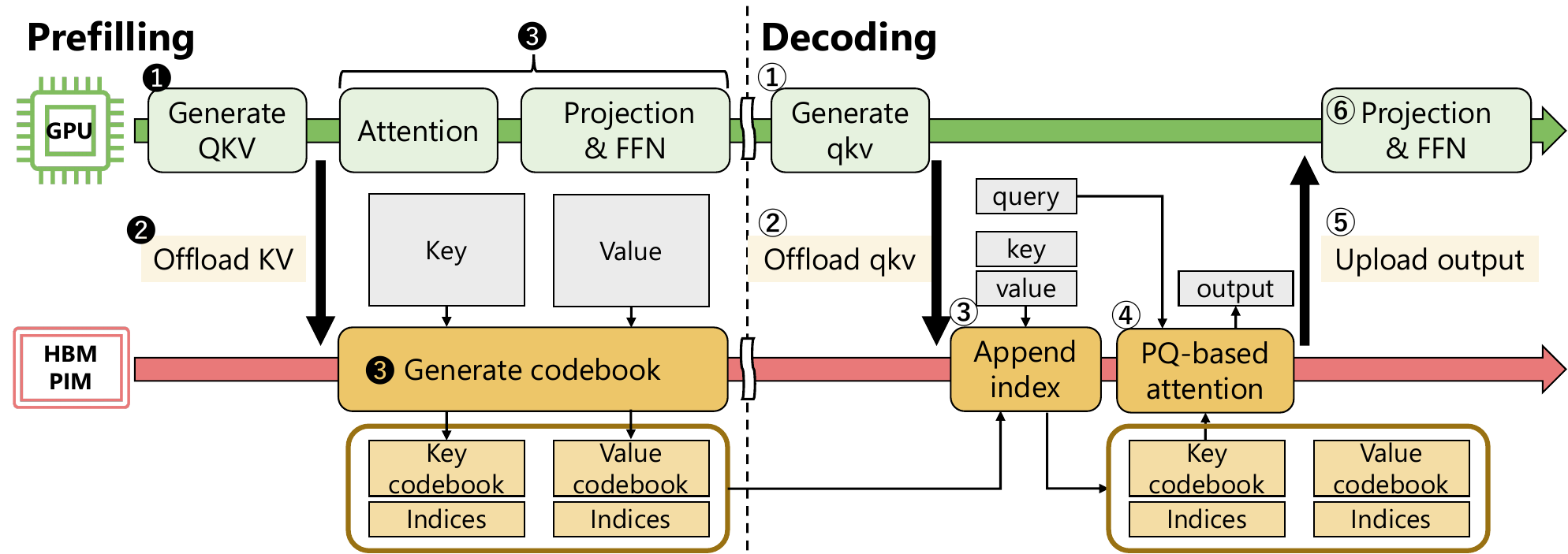}
        \caption{
        \AQPIM execution flow during prefilling and decoding with GPU and HBM-PIM. 
        }
        \label{fig:overview}
    \end{subfigure}
    \hspace{1mm}
    \begin{subfigure}{0.32\linewidth}
        \centering
        \includegraphics[height=4.3cm]{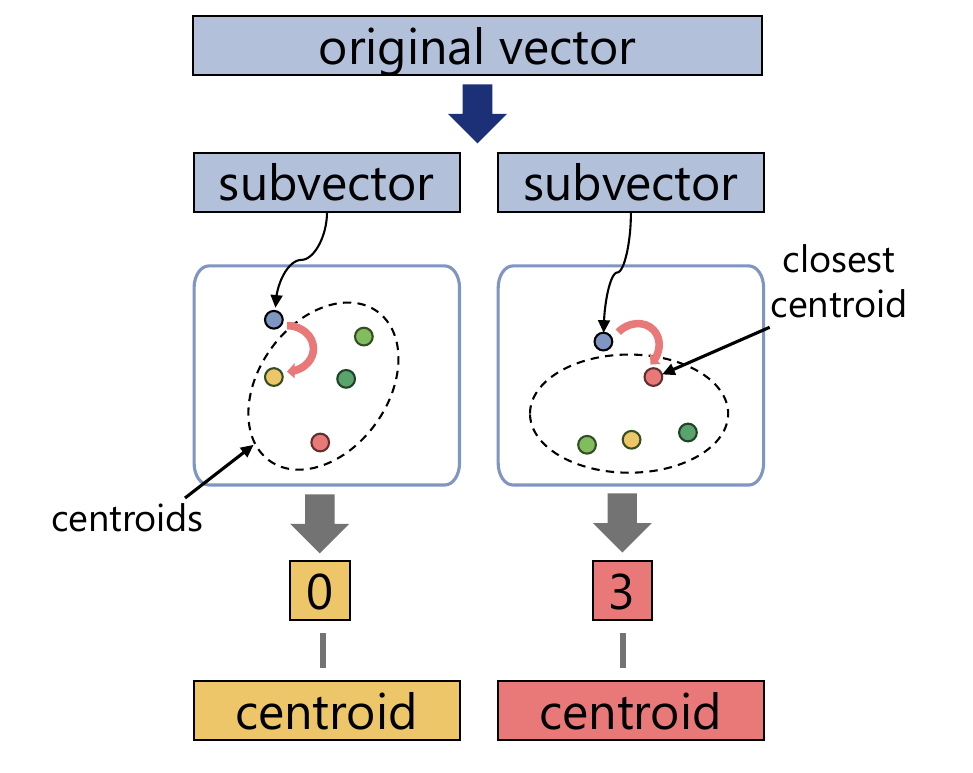}
        \caption{
        PQ applies clustering-based quantization.
        }
        \label{fig:pq}
    \end{subfigure}
    \vspace{-1mm}
    \caption{Overview of \AQPIM and Product Quantization (PQ).}
    \vspace{-2mm}
\end{figure*}

\subsection{Overview}
\label{ssec:overview}
This paper proposes \AQPIM, an activation quantization framework utilizing PIM for efficient activation handling and attention computation in large-scale models. Leveraging PIM's high internal bandwidth, \AQPIM employs an online, context-aware clustering-based quantization to compress activations. Furthermore, \AQPIM uses the resulting \emph{codebooks}, i.e., data structures originally used to reproduce vectors, directly for GEMV. This enables attention computation directly on the compressed data (without decompression) and repetitive reuse of the partial results, significantly reducing both memory footprint and computational overhead within the PIM architecture.

\cref{fig:overview} provides an overview of \AQPIM. Similar to prior work~\cite{park2024attacc!, heo2024neupims}, \AQPIM leverages both the high computational power of GPUs and the high intra-memory bandwidth of PIMs.
During the prefilling, GPU generates the QKV matrices \ding{182} and offloads KV to PIM \ding{183}. Then, GPU computes attention and processes projection and feedforward network (FFN)  \ding{184}. 
Meanwhile, PIM generates the \emph{codebooks} and mapping indices with key and value clustering and compression \ding{184} in parallel with the GPU execution. 
During the decoding, GPU generates the qkv vectors (hereafter, vectors are expressed with lower case) \ding{192} and sends them to PIM \ding{193}. 
Subsequently, PIM appends their indices \ding{194} and computes the attention output using the compressed format \ding{195}. Finally,  attention output is transferred back to the GPU \ding{196}, followed by GPU's processing projection and feedforward operations \ding{197}. 

The sequential GPU-PIM processing during the decoding phase may cause GPU idling, especially when the context gets long.
This is mitigated by sequence-by-sequence pipelining, where GPU generates query, key, and value vectors for each sequence and immediately offloads them to the PIM, while it proceeds to process the next sequence. 

\subsection{Product Quantization}
\label{ssec:product_quantization}
As motivated in Section~\ref{ssec:motivation}, \AQPIM is designed to: (a) leverage context-dependent similarity and locality through its quantization scheme, (b) achieve a balanced trade-off between compression efficiency and PIM-suitable bandwidth, (c) effectively utilize the localized memory scope of PIM, and (d) minimize both memory footprint and computation within the PIM architecture. To this end, we adopt Product Quantization (PQ) as our baseline quantization technique.

PQ is a vector quantization technique widely recognized in the approximate nearest neighbor search (ANNS) for its capacity to significantly compress high-dimensional vector data while preserving locality and similarity. 
As illustrated in~\cref{fig:pq}, PQ has two fundamental characteristics: (1) \textit{vector splitting} and (2) \textit{clustering-based quantization}.
(1) \textit{Vector Splitting} decomposes high-dimensional vectors into smaller subvectors. This allows for parallel processing across subvector groups, effectively utilizing PIM's high parallelism and localized memory scope, while improving expressibility by combining multiple subvector spaces to reconstruct a vector.
(2) \textit{Clustering-based quantization} significantly reduces quantization error by exploiting similarity and locality in data distribution. PQ typically employs k-means clustering, which partitions vectors into $k$ clusters based on the Euclidean distance. Since distance calculation for each vector is independent, this approach aligns well with the parallel processing capabilities of PIM.

Clustering has been used to \emph{identify} important tokens in offloading-based sparse attention approaches like PQCache~\cite{zhang2024pqcacheproductquantizationbasedkvcache} and Squeezed Attention~\cite{hooper2025squeezedattentionacceleratinglong}, where a \emph{full exact copy of KVs is kept at and fed from CPU}. Importantly, \AQPIM \emph{directly} uses PQ as a quantization method and KV source in PIM without the full KV copy. This eliminates the need for CPU offloading and bandwidth for KV transfers, while we observe that the naive adoption of PQ as a KV source results in a non-trivial accuracy drop. This will be addressed by our algorithmic techniques introduced in Sections~\ref{ssec:weighted} and \ref{ssec:presort}. 

\textbf{PQ for KV Cache Quantization}: 
PQ's codebook generation can be a significant bottleneck when applying PQ for KV cache quantization during inference. To overcome this issue, we leverage underutilized PIM resources during the prefilling stage. \cref{fig:overview} shows parallel processing of GPU and PIM. PIM generates the codebooks concurrently with GPU computation during the prefilling stage.

\begin{figure}[tb]
    \centering
    \includegraphics[width=\linewidth]{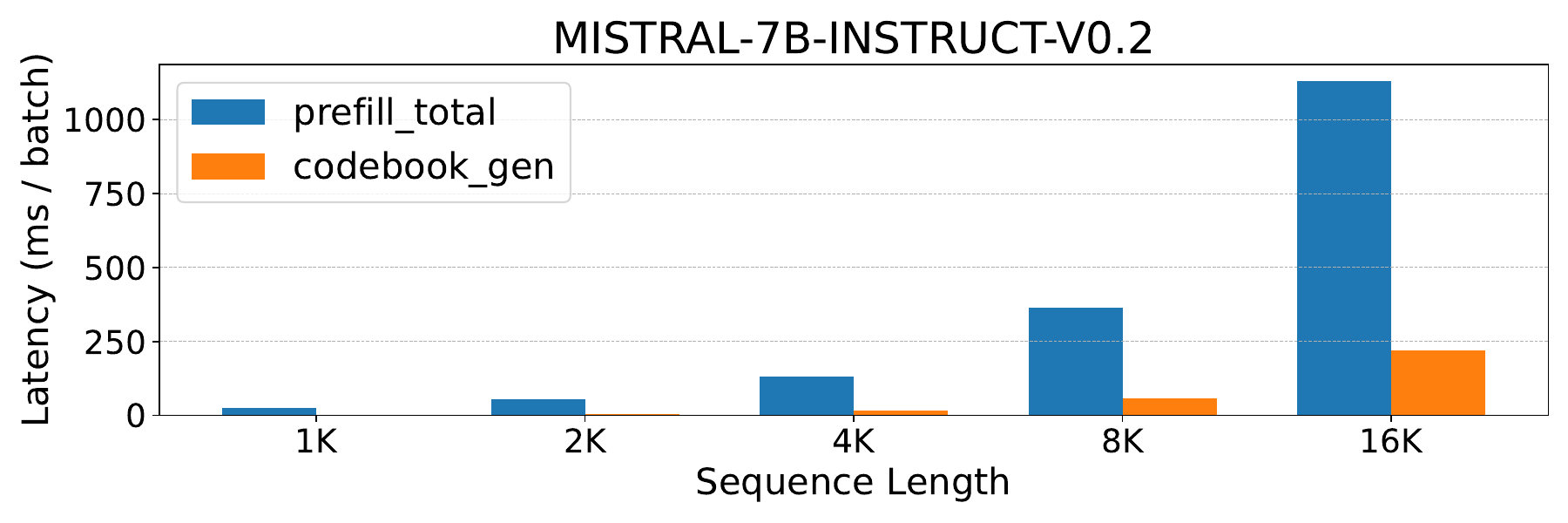}
    \caption{\RBtext{The latency comparison of the prefilling and the clustering process in 128 head-dimensional space.}}
    \label{fig:codebook_gen_overhead}
    \vspace{-3mm}
\end{figure}

To keep up with the GPU's throughput, codebook generation must be completed before the GPU offloads KV of the subsequent layer. While standard k-means iterates cluster reassignment until convergence, our experiments demonstrate that just \textit{four} iterations converge to a stable state and yield comparable accuracy, effectively hiding the clustering process behind the GPU's computations.
\RBtext{
\RB{B1}\label{RB:B1}
The clustering overhead can be completely hidden regardless of sequence length, 
as shown in \cref{fig:codebook_gen_overhead}.
Given a vector PEs, while the latency for attention scales with $N^2$, clustering scales with $n_{\text{centroids}}N$, where $n_{\text{centroids}}$ is a constant and $n_{\text{centroids}} \ll N$. 
Furthermore, peak memory usage is minimized by layer-wise codebook generation, enabling sequential compression of KV cache.
}

\begin{figure}
    \centering
    \includegraphics[width=\linewidth]{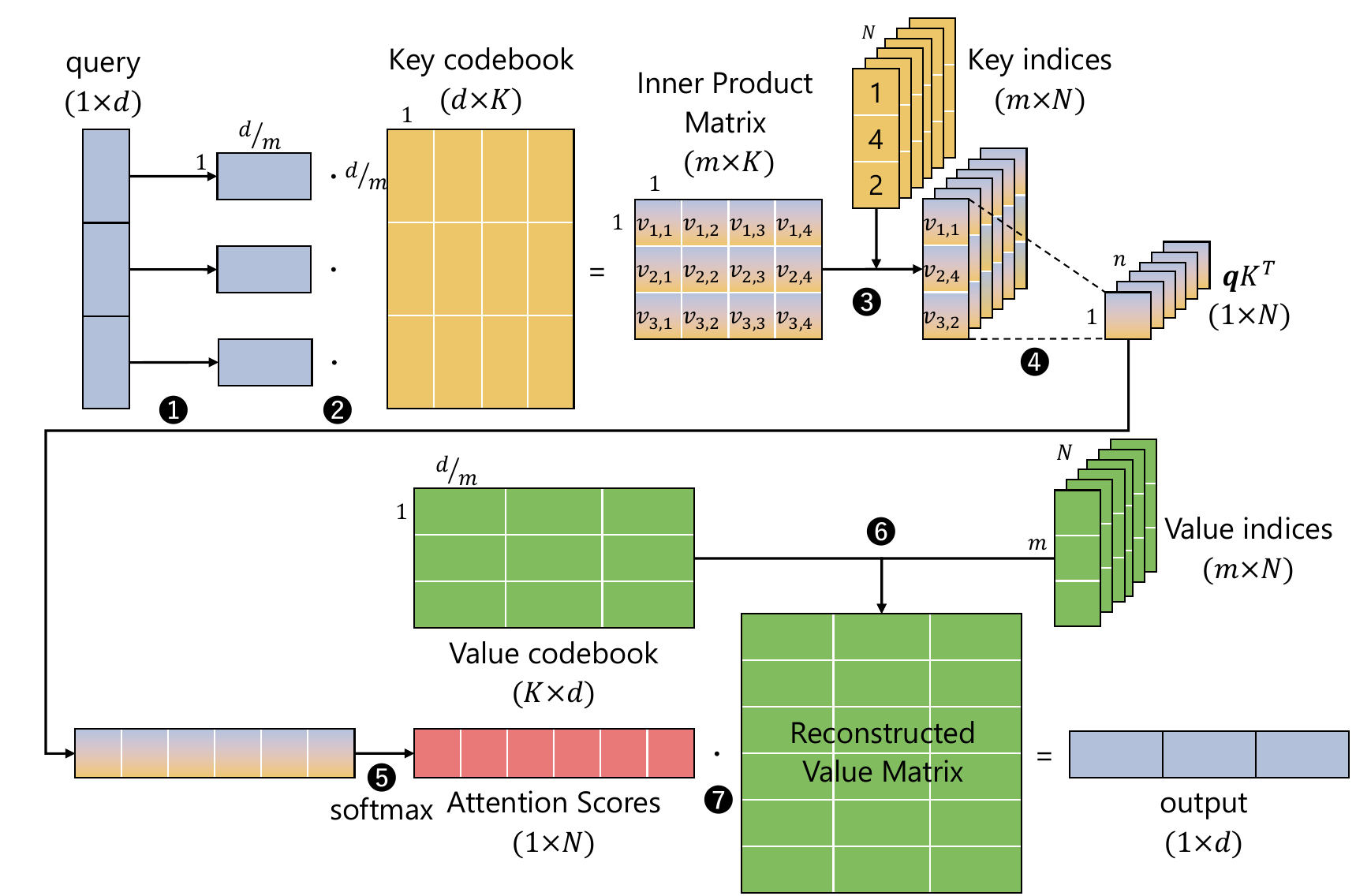}
    \caption{
    Computation flow of PQ-based attention. Matrix multiplications are simplified by inner product matrix lookup and summation.
    }
    \vspace{-3mm}
    \label{fig:pq-based_attention}
\end{figure}

\textbf{PQ for Efficient Attention Computation}: 
PQ significantly optimizes attention computation by directly leveraging codebook representations. In PQ, the key cache is decomposed into a key codebook and a set of key indices. The key codebook contains centroids generated during the prefilling stage, and the key indices indicate the centroid assignment for each original token. Since our goal is to compute the inner product of the query vector and the key matrix, the reconstruction of the full key matrix can be skipped. \cref{fig:pq-based_attention} illustrates this skip method.
The query vector is first divided into $m$ subvectors \ding{182}. Subsequently, each subvector is multiplied with its corresponding codebook's submatrix, collectively forming an inner product matrix \ding{183}. The key indices, indicating centroid assignments, are then used to lookup values in the inner product matrix \ding{184}. The retrieved values are summed along the vector splitting axis \ding{185}, which produces an approximation of the inner product $\boldsymbol{q}K^T$. This sequence of operations \ding{182}–\ding{185} dramatically reduces the computational cost of obtaining $\boldsymbol{q}K^T$ by avoiding the explicit multiplication between the query and the full key matrix.
Subsequently, the softmax function is applied to $\boldsymbol{q}K^T$ to produce the attention scores \ding{186}. The value matrix is reconstructed using the value codebook and value indices \ding{187}. Finally, the output vector is computed as the inner product of the attention scores and the reconstructed value matrix \ding{188}.
This method achieves substantial computational savings by replacing large-scale matrix multiplications with efficient, localized codebook lookups and summations. Our PQ-based attention mechanism is orthogonal to recent techniques such as Grouped-Query Attention (GQA) and Multiple-Query Attention (MQA).


\textbf{Mitigating Random Access for Efficient Lookup}:
A naive implementation of the inner product matrix lookup in PQ-based attention generates irregular memory accesses that lead to frequent DRAM row activations.

To address this issue, we propose a \emph{page-aware windowed clustering} method, co-designed with intra-row indirection support in \AQPIM introduced later in \cref{ssec:row_buffer_retrieval}.
The core idea is to restrict the indirect access to happen within a single DRAM row in a bank by co-locating all related inner product values. 

\begin{figure}[tb]
    \centering
    \includegraphics[width=\linewidth]{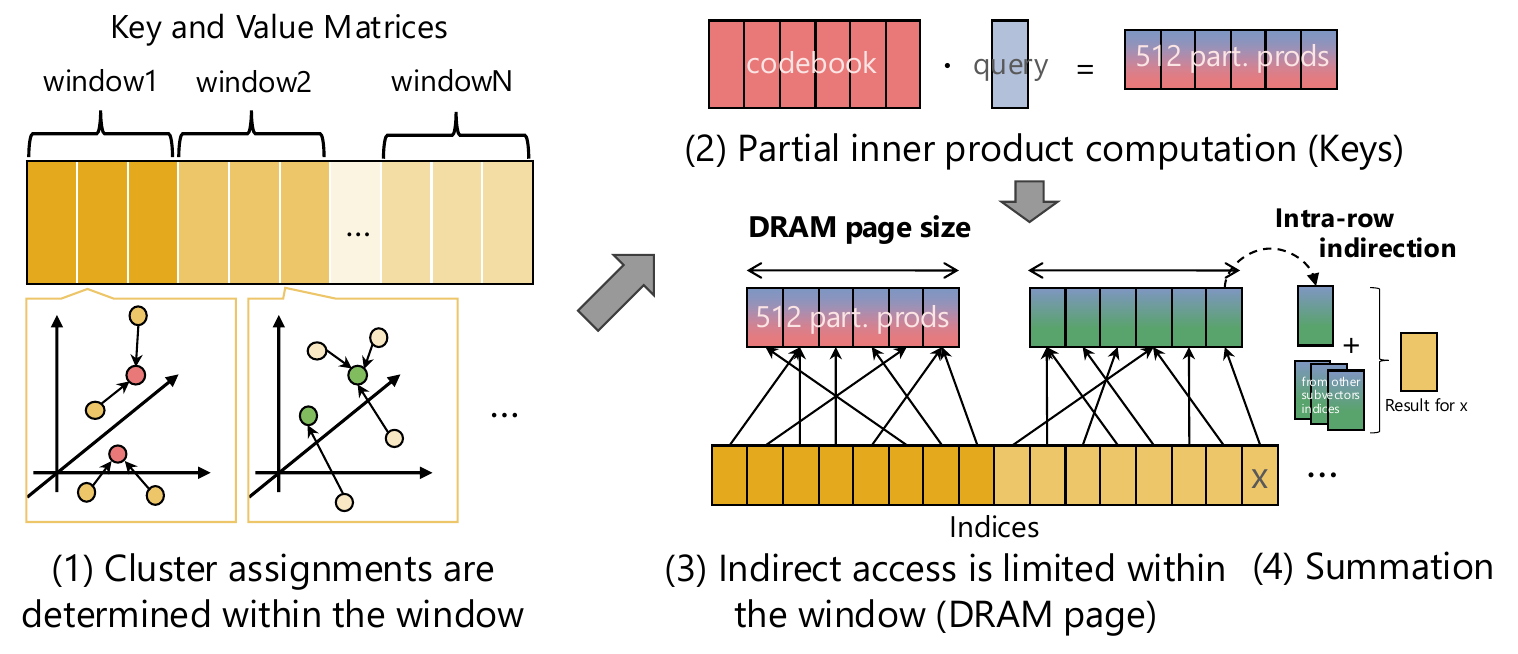}
    \vspace{-6mm}
    \caption{\RBtext{Page-aware windowed clustering.}}
    \label{fig:pawc}
    \vspace{-3mm}
\end{figure}

For example, HBM-PIM architecture has 1KB row buffers in each bank, which stores 512 inner product values (in FP16 format), and we use that many centroids for a given context window so that indirection only happens within a page.
Although a single window that maps the entire sequence to 512 centroids suffices for most long-context scenarios, it can be extended for more centroids. To do this, \RBtext{as shown in \cref{fig:pawc}\RB{B2}\label{RB:B2} (1) a sequence is divided into multiple windows, and as the window advances, the previous centroids are copied to a new page and subsequently updated for the window. 
Then, (2) partial inner product is computed, and (3) intra-row indirection performs lookups within a DRAM page.  
Since indirect access within each window is fully contained within a single row, it greatly minimizes the number of row activations, reducing it to as few as the number of windows. 
The partial results are then (4) summed up with partial products from different subvectors. 
}

This approach is also extended for the value codebook, while we may loop over the value indices multiple times to accommodate the larger tensor dimensions. 


    
            
            

\subsection{Weighted Codebook Generation}
\label{ssec:weighted}
Despite PQ offering potential benefits of acceleration of attention computation, just applying standard PQ to attention layers leads to substantial accuracy loss. Consequently, prior work PQCache~\cite{zhang2024pqcacheproductquantizationbasedkvcache} limits its use to identifying important tokens and retrieves their original KVs from CPU memory. We hypothesize that this accuracy degradation stems from PQ's inherent inability to account for the varying levels of importance among different tokens. Previous studies~\cite{xiao2024efficient, zhang2023ho} have shown that certain tokens consistently receive high attention scores. These critical tokens play a crucial role in preserving model accuracy, but the conventional PQ treats all tokens equally during quantization.

To address this problem, we propose importance-weighted k-means clustering, ensuring that tokens with higher attention scores are quantized with fewer quantization errors.
We modify the k-means clustering process by incorporating attention score-based weighting, giving higher priority to tokens with greater impact on the model accuracy. 
First, we compute a weight vector $\boldsymbol{w}\in\mathbb{R}^{N}$ from the attention score matrix $\mathbf{S}\in\mathbb{R}^{N\times N}$.
Each weight is the sum of attention scores received from the last $t$ tokens in the sequence:
\begin{equation}\label{eq:weight}
    \boldsymbol{w} = \operatorname{sum}(\mathbf{S}[-t:, :], \operatorname{axis}=0),
\end{equation}
where $t$ is a tunable window size.
The algorithm then iteratively minimizes a weighted objective function
, which is the total weighted squared Euclidean distance between each token $\boldsymbol{x}_n$ and its assigned cluster centroid $\boldsymbol{\mu}_k$.
In each iteration, tokens are assigned to their closest centroid, and centroids are subsequently updated using a weighted average of their members:
\begin{equation}
\label{eq:centroid_calculation}
    \boldsymbol{\mu}_k=\frac{\sum_{n\in C_k}w_n\boldsymbol{x}_n}{\sum_{n\in C_k}w_n},
\end{equation}
where $C_k$ is the set of indices for the tokens assigned to the $k$-th cluster.
By introducing these weights, the centroids are more influenced by tokens exhibiting high attention scores, thereby greatly reducing quantization error for these critical tokens. As a result, this approach improves accuracy retention while preserving the benefits of PQ compression.


The weights $\boldsymbol{w}$ defined in \cref{eq:weight} are computed on the GPU during the prefilling phase.
Since the attention score $\mathbf{S}$ is used in both attention and weights computations, the additional computational overhead is minimal and aligns with FlashAttention\cite{dao2022flashattention}.

\subsection{Optimization of Vector Splitting}
\label{ssec:presort}
Standard PQ splits vectors without considering inter-channel similarity, which often leads to higher quantization errors. To address this, we introduce a channel sorting preprocessing step. By grouping highly correlated channels together before partitioning, this approach creates more cohesive subvectors, thereby minimizing quantization error while maintaining the original codebook size.

Our sorting method groups channels based on cosine similarity.
The process begins by randomly selecting a reference channel.
The cosine similarity is then computed between this reference channel and all other channels.
Based on the results, the top-$k$ most similar channels are greedily selected to form a group.
These steps are repeated $m$ times, where $m$ is the number of subvectors, until every channel has been assigned to one of the groups.
As a result, channel vectors within each group exhibit high mutual cosine similarity.
\RBtext{Compared to coupling\RB{A2}\label{RB:A2} contiguous channels for quantization~\cite{zhang2024kvcache}, pre-sorting helps increase intra-group affinity, reducing quantization error.}


This channel sorting operation can be seamlessly integrated into the projection matrices, effectively hiding any associated overhead. 
Following the approach of~\cite{duanmu2024skvq}, we introduce static sorting matrices, $P_k$ and $P_v$ for the key and value vectors.
The sorting matrices can be absorbed into the projection matrices. Specifically, they can be incorporated as $W_q'=W_qP_k$, $W_k'=W_kP_k$, $W_v'=W_vP_v$, and $W_o=W_oP_v^T$. Moreover, these sorting matrices are generated offline using a calibration dataset, such as Wikitext-2-v1~\cite{merity2017pointer}, thereby avoiding additional runtime overhead during inference.

\subsection{System Architecture}
\label{ssec:system_architecture}
We consider a hardware design based on HBM~\cite{jedec2022hbm3} integrated with PIM. HBM provides high memory bandwidth thanks to its 3D-stacked DRAM dies, which are interconnected via through-silicon vias (TSVs). This 3D architecture also enables high energy efficiency since data can be transferred over much shorter distances. HBM has been adopted in recent GPUs such as NVIDIA's H100~\cite{nvidia2024h100}, making it a suitable candidate for integration with PIM.

\begin{figure}[tb]
    \centering
    \includegraphics[width=\linewidth]{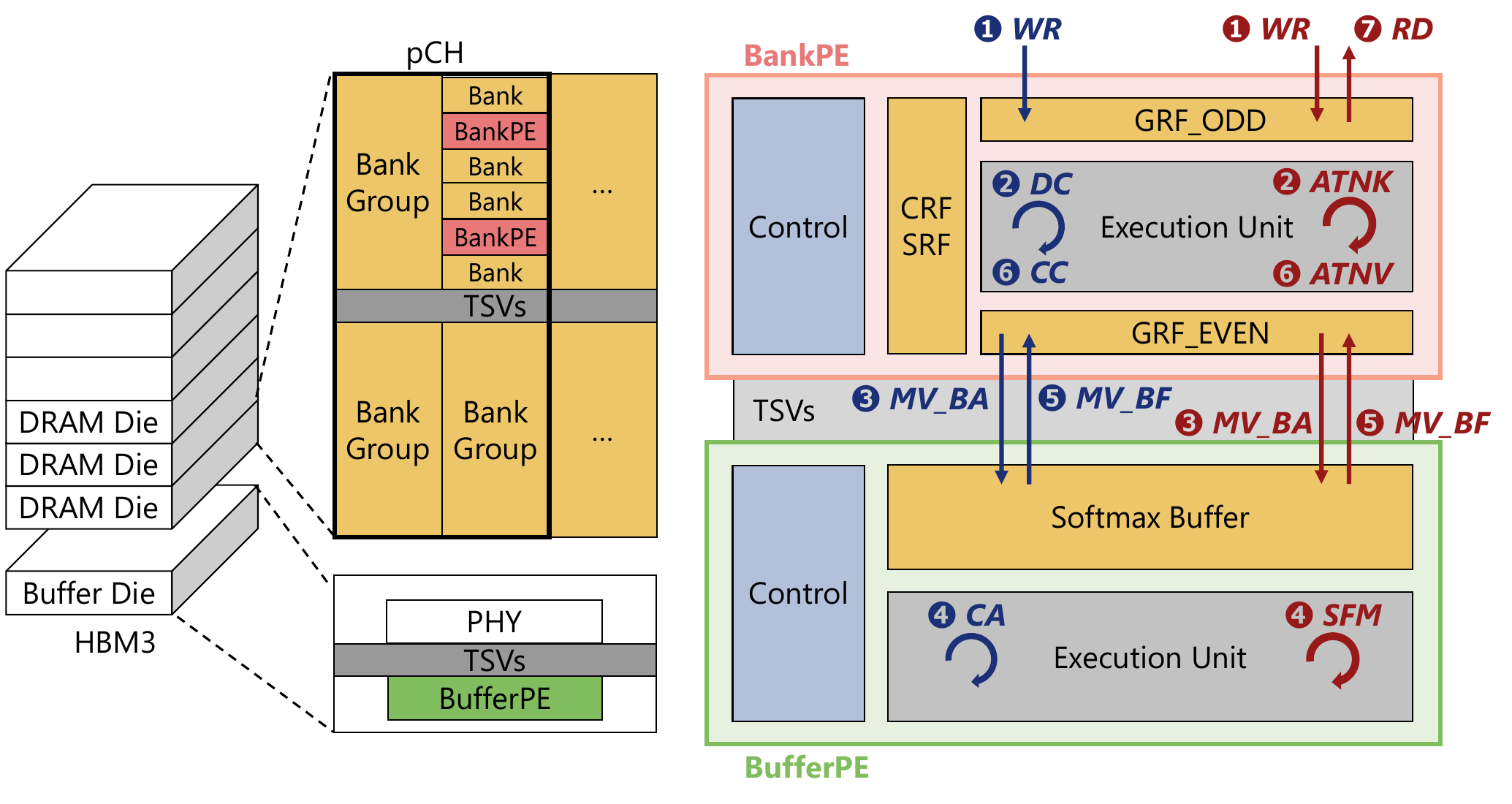}
    \vspace{-5mm}
    \caption{
    \AQPIM architecture and dataflow.
    }
    \label{fig:hbm-pim_overview}
\end{figure}

\textbf{Execution Unit}: A key challenge in HBM-PIM design is the placement of PE, which significantly impacts throughput and energy efficiency. To address this issue, AttAcc!~\cite{park2024attacc!} explores PE placement in terms of peak power, throughput, energy consumption, and area overhead. Building on this analysis, we introduce two PE architectures: BankPE and BufferPE, as illustrated in~\cref{fig:hbm-pim_overview}. BankPE is placed adjacent to the DRAM banks, leveraging the high internal bandwidth while facing strict area constraints. In contrast, BufferPE is located in the buffer die of HBM. Although it provides lower bandwidth compared to BankPE, it benefits from fewer area constraints. Moreover, BufferPE is particularly advantageous for data-intensive operations that involve inter-bank data movement, which would otherwise introduce bottlenecks in BankPEs.
Importantly, we design the microarchitectures for BankPE and BufferPE based on their individual strengths and limitations.
While we utilize the architectural modules based on AttAcc!, we optimize them to be well-suited for our \AQPIM algorithms.

We identify the necessary computation unit for PQ and attention, as shown in~\cref{tb:operation_breakdown}. Distance calculation (DC), centroid calculation (CC), and attention computation (ATNK, ATNV) are not data-intensive. This motivates placing their corresponding computation units in BankPE. Moreover, since some units overlap, frequently used operations can be efficiently assigned to BankPE, minimizing its area overhead. From this perspective, ADD, MUL, and SUM units are placed in BankPE. In contrast, data-intensive operations, such as cluster assignment (CA) and softmax-related operations (SFM), are assigned to BufferPE to reduce inefficient inter-bank data transfers. This placement is also suitable because the DIV and EXP calculators consume relatively large chip area. 
Note that \AQPIM does not introduce any specialized computation units or ALUs for quantization to area constraint BankPEs and only uses existing FP16 MAC units.

\begin{table}[tb]
    \centering
    \caption{
    List of \AQPIM operations. 
    }
    \vspace{-2mm}
    \label{tb:operation_breakdown}
    {\footnotesize
    \begin{tabular}{c|c|c}
        \bhline{1pt}
        Process & Place & Necessary Unit \\
         \hline
        Distance Calculation (DC) & BankPE & ADD, MUL, SUM \\
        Cluster Assignment (CA) & BufferPE & MIN \\
        Centroid Calculation (CC) & Both & MUL, SUM, DIV \\
        Attention (ATNK, ATNV) & BankPE & MUL, SUM \\
        Softmax (SFM) & BufferPE & ADD, SUM, MAX, DIV, EXP \\
        \bhline{1pt}
    \end{tabular}
    }
\end{table}

\textbf{Dataflow}: \cref{fig:hbm-pim_overview} also illustrates the data flow for both codebook generation (blue arrows) and attention computation (red arrows). 
Codebook generation begins with receiving KV matrices from the GPU, which are then distributed to each BankPE \ding{182}. Each BankPE performs distance calculation (DC) \ding{183}, and the results are transmitted to the BufferPE \ding{184}. The BufferPE then determines cluster assignments (CA) \ding{185} and returns the assignment results to their respective banks \ding{186}. Based on these assignments, each BankPE performs centroid calculation (CC) \ding{187}. The BankPE computes the numerator (a weighted sum of vectors), while the BufferPE computes the reciprocal of the denominator (a sum of weights), as presented in~\cref{eq:centroid_calculation}. The final division is thus reduced to a single multiplication at the BankPE. Steps \ding{183}-\ding{187} are iteratively repeated until the codebook is ultimately generated.

The attention computation begins with the query vector being received \textcolor{purple}{\ding{182}}. Each BankPE performs multiplication between the query and key codebook (ATNK) \textcolor{purple}{\ding{183}}, and transfers the results to the BufferPE \textcolor{purple}{\ding{184}}. The BufferPE then computes the softmax function (SFM) \textcolor{purple}{\ding{185}} and returns the results to each BankPE \textcolor{purple}{\ding{186}}. Finally, each BankPE performs the final attention computation (ATNV) \textcolor{purple}{\ding{187}}, sending the attention output to the GPU \textcolor{purple}{\ding{188}}.

\textbf{Commands}: Several dedicated PIM commands are introduced on top of the AttAcc! design to control PQ-related processes. 
\texttt{PIM\_SET\_CONFIG} broadcasts the PQ configuration, including parameters such as the number of subvectors and the number of centroids. 
\texttt{PIM\_MAC\_AB} executes MAC operations in all banks.
\texttt{PIM\_SFM} executes softmax-related operations within BufferPE.
\texttt{PIM\_RET} executes row buffer retrieval explained in \cref{ssec:row_buffer_retrieval}.
In addition to computational commands, we use several data movement commands. \texttt{PIM\_MV\_BA} command moves data from BankPE to BufferPE, and \texttt{PIM\_MV\_BF} command transfers data from BufferPE back to BankPE. To manage I/O,  \texttt{PIM\_RD} command reads the final results of attention computation from the bank, and \texttt{PIM\_WR} writes input data to the bank as the initial step of the processing. Finally, to ensure proper DRAM operation, we include system-level commands such as \texttt{PIM\_ACT\_AB} command, which activates DRAM rows in all banks. 
Although these commands are not yet implemented on HBM-PIM, they are issued through the standard HBM command path in the same manner as conventional DRAM commands.

\subsection{Intra-Row Indirection}
\label{ssec:row_buffer_retrieval}
\begin{figure}
    \centering
    \includegraphics[width=\linewidth]{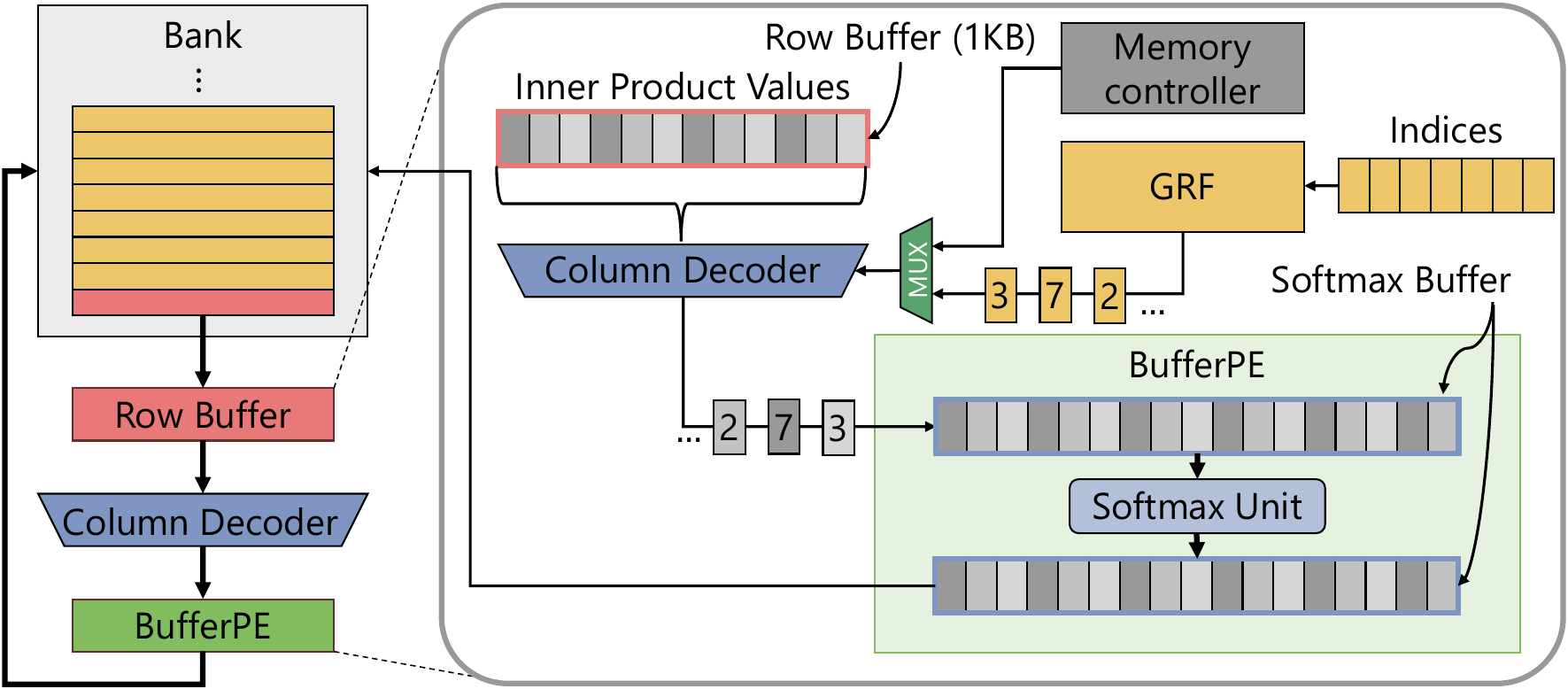}
    \caption{Intra-row indirection for efficient random access. 
    }
    \label{fig:row_buffer_retireval}
    \vspace{-3mm}
\end{figure}

Since clustering assigns arbitrary centroids, the following lookup operations during attention computation involve random access to the inner product table.
To efficiently manage these irregular memory access patterns, we propose the intra-row indirection architecture, as illustrated in \cref{fig:row_buffer_retireval}.
This mechanism operates as follows: 
First, the row storing the target (inner product) values is activated, and transfer the data into a row buffer.
Then, the lookup indices stored in a general-purpose register file (GRF) are redirected to the column decoder.
The column decoder then outputs the corresponding values, which are streamed through the existing datapaths to the buffer die or GRF.
The BufferPE executes the subsequent softmax operation and transfers the results back to the banks.

Importantly, only a single row activation is necessary as long as the memory address scope represented by indices fits entirely within a row buffer, which is guaranteed by page-aware windowed clustering.
Moreover, this mechanism incurs no significant additional area overhead, making it well-suited for BankPE, which often faces severe area constraints.




\subsection{Data Mapping Strategy}
\label{ssec:data_mapping_strategy}
\begin{figure}[tb]
    \centering
    \includegraphics[width=\linewidth]{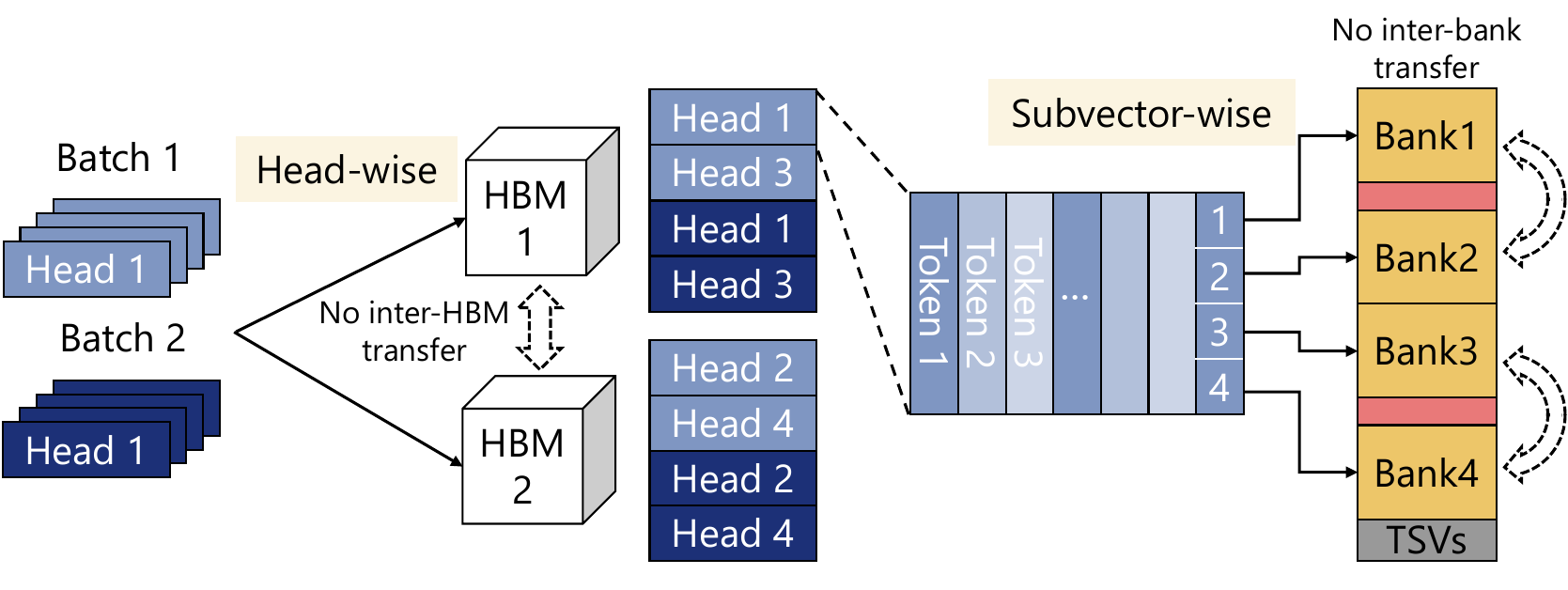}
    \vspace{-5mm}
    \caption{
    Data mapping strategy. Each head is assigned to a separate HBM, and each subvector set is mapped to an individual memory bank.
    }
    \vspace{-3mm}
    \label{fig:mapping_strategy}
\end{figure}

We design an efficient data mapping strategy to maximize PIM utilization, as illustrated in~\cref{fig:mapping_strategy}. Attention computation is inherently parallel due to mechanisms such as MHA and GQA. In addtion, PQ's codebook is applied per head. 
From this, each attention head is mapped to a separate HBM. \RBtext{This\RB{D5}\label{RB:D5} head-wise mapping effectively eliminates unnecessary data transfers between HBMs, ensuring a high utilization even when multiple heads share a stack.} Moreover, within the \AQPIM framework, each head is further split into multiple subvectors. To maximize BankPEs utilization, each subvector should be assigned to a different bank. 
\RBtext{This implies that\RB{C4}\label{RB:C4} PIM utilization is maximized if $N_{\text{subvecs}} > \frac{N_{\text{banks}}}{N_{\text{heads}} \times N_{\text{batches}}}$. It is easy to satisfy this condition with practical LLM setups.}
This bank-wise data mapping is realized by simply incrementing the address because
\AQPIM employs an address mapping scheme that utilizes the lower bits of the memory address for bank selection.

\subsection{Memory Allocation}
To use memory space effectively and minimize unnecessary memory remapping, the memory region in each bank must be handled properly.
We utilize a simple and effective memory allocating strategy.
The memory region is allocated before the prefilling phase. The codebook region is fixed and constant, and the buffer region used during prefilling is also fixed and reused for each layer.
PQ indices are allocated layer by layer at the page granularity. A fixed-size PQ index region for prefilling phase is allocated between the codebook and buffer regions and overwrites the buffer region thereafter during decoding. 
These regions need not be dynamically reclaimed, but released at once after completion. 

%% file: sections/4_evaluation.tex
\section{Evaluation}
\label{sec:evaluation}

\subsection{Experimental Setup}
\label{ssec:experimental_setup}
\textbf{Models}: 
We conduct experiments using two open-source LLMs: Mistral-7B-Instruct-v0.2~\cite{Jiang2023mistral7b} and Llama-3.2-3B-Instruct~\cite{grattafiori2024llama3herdmodels}. These models have long context window sizes of 32K and 128K, respectively. We use bfloat16 for both models, which is a common format for LLM inference.

\textbf{Tasks}: 
We evaluate \AQPIM using LongBench~\cite{Bai2024longbench}, a widely used benchmark for long-context LLM inference. LongBench includes diverse task categories such as single- and multi-document question answering, summarization, few-shot learning, synthetic tasks, and code completion.
To evaluate overall accuracy trends, we select one representative task from each of these categories for our experiments.

\textbf{Baselines}: 
We compare \AQPIM with 
SnapKV~\cite{li2024snapkv},  PQCache~\cite{zhang2024pqcacheproductquantizationbasedkvcache}, and SKVQ~\cite{duanmu2024skvq}.
SnapKV is also a state-of-the-art sparse attention method that dynamically selects tokens during inference, enabling efficient long-context processing. 
PQCache is an offloading method that uses PQ to identify important tokens. It offloads KV cache to CPU memory and retrieves a small number of tokens based on a maximum inner product search with PQ. 
SKVQ is a state-of-the-art quantization approach that reorders channels to minimize the quantization error within each quantization group.

\textbf{Hyperparameters}: 
Based on the configurations of PQCache and SKVQ, we retain the first 8 tokens with full precision, which is well known as sink tokens. We adopt the same approach in \AQPIM.
In addition, similar to other methods, \AQPIM preserves the most recent 32 tokens, referred to as sliding window tokens, with full precision.
We also use these 32 ($=t$) tokens to calculate weight $\boldsymbol{w}$ defined in \cref{eq:weight}.
\AQPIM also has two key hyperparameters: the number of subvectors and the number of centroids. 
First, to determine the optimal number of subvectors, we conduct experiments on a subset of LongBench with Mistral-7B-Instruct-v0.2. 
In these experiments, we vary the number of subvectors $m$ while keeping the number of centroids fixed at 512. As shown in~\cref{tb:num_of_sub-vectors}, using 32 subvectors achieves the best balance.
Subsequently, we varied the number of centroids, observing that accuracy saturated at 512 centroids, as shown in \cref{tb:num_of_centroids}. This number of centroids (512) is also well-suited for intra-row indirection, explained in \cref{ssec:row_buffer_retrieval}.

\RBtext{
\label{RB:D6}
\textbf{Online codebook update}: 
We tried OnlinePQ~\cite{xu2018onlineproductquantization},\RB{D6} which progressively updates centroids at each decoding step. However, it is observed to have little impact on accuracy, even on LongBench. Moreover, in some cases, it negatively affected accuracy; for example, the average score of LongBench tasks is 0.36 points lower than that of the non-OnlinePQ configuration. Therefore, we omit OnlinePQ from our experiments.
}


\begin{table}[tb]
    \centering
    \caption{Accuracy comparison across different number of subvectors $m$.}
    \label{tb:num_of_sub-vectors}
    \begin{tabular}{c|cccccc}
        \bhline{1pt}
        Configuration & m=2 & m=4 & m=8 & m=16 & m=32 & m=64 \\
        \hline
        NarrativeQA & 21.45 & 20.58 & 21.36 & 22.09 & 22.59 & 21.81 \\
        HotpotQA & 31.35 & 32.31 & 35.71 & 37.40 & 37.83 & 37.85 \\
        GovReport & 21.05 & 21.42 & 22.49 & 25.91 & 29.88 & 30.73 \\
        TREC & 51.00 & 57.00 & 65.50 & 70.00 & 71.00 & 71.00 \\
        PRetrieval & 86.35 & 87.13 & 88.81 & 88.19 & 87.69 & 86.85 \\
        LCC & 53.83 & 54.89 & 55.03 & 55.45 & 55.33 & 55.23 \\
        \hline
        Average & 44.17	& 45.56	& 48.15 & 49.84	& \textbf{50.72} & 50.58 \\
        \bhline{1pt}
    \end{tabular}
\end{table}

\begin{table}[tb]
    \centering
    \caption{Accuracy comparison across different number of centroids $K$.}
    \label{tb:num_of_centroids}
    \begin{tabular}{c|ccccc}
        \bhline{1pt}
        Configuration & K=64 & K=128 & K=256 & K=512 & K=1024 \\
        \hline
        NarrativeQA & 23.25 & 23.09 & 21.37 & 22.59 & 21.91 \\
        HotpotQA & 37.38 & 38.08 & 38.00 & 37.83 & 38.13 \\
        GovReport & 22.99 & 25.49 & 28.53 & 29.88 & 30.74 \\
        TREC & 69.00 & 70.50 & 71.00 & 71.00 & 71.00 \\
        PRetrieval & 81.56 & 88.17 & 87.85 & 87.69 & 87.36 \\
        LCC & 54.08 & 54.66 & 55.17 & 55.33 & 55.20 \\
        \hline
        Average & 48.04 & 50.00 & 50.32 & \textbf{50.72} & \textbf{50.72} \\
        \bhline{1pt}
    \end{tabular}
\end{table}

\begin{figure*}
    \begin{subfigure}{0.50\linewidth}
        \centering
        \includegraphics[width=\linewidth]{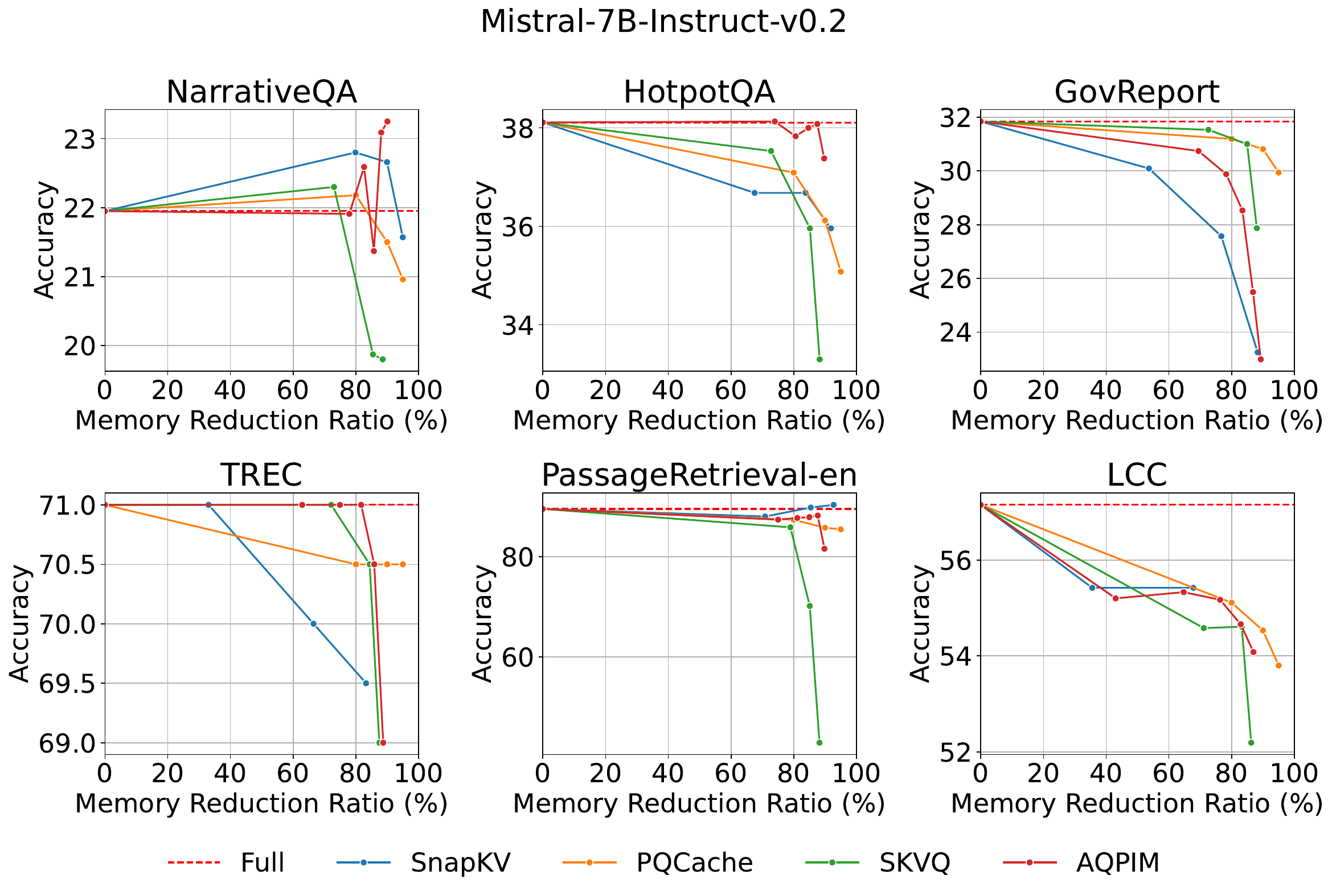}
    \end{subfigure}
    \begin{subfigure}{0.50\linewidth}
        \centering
        \includegraphics[width=\linewidth]{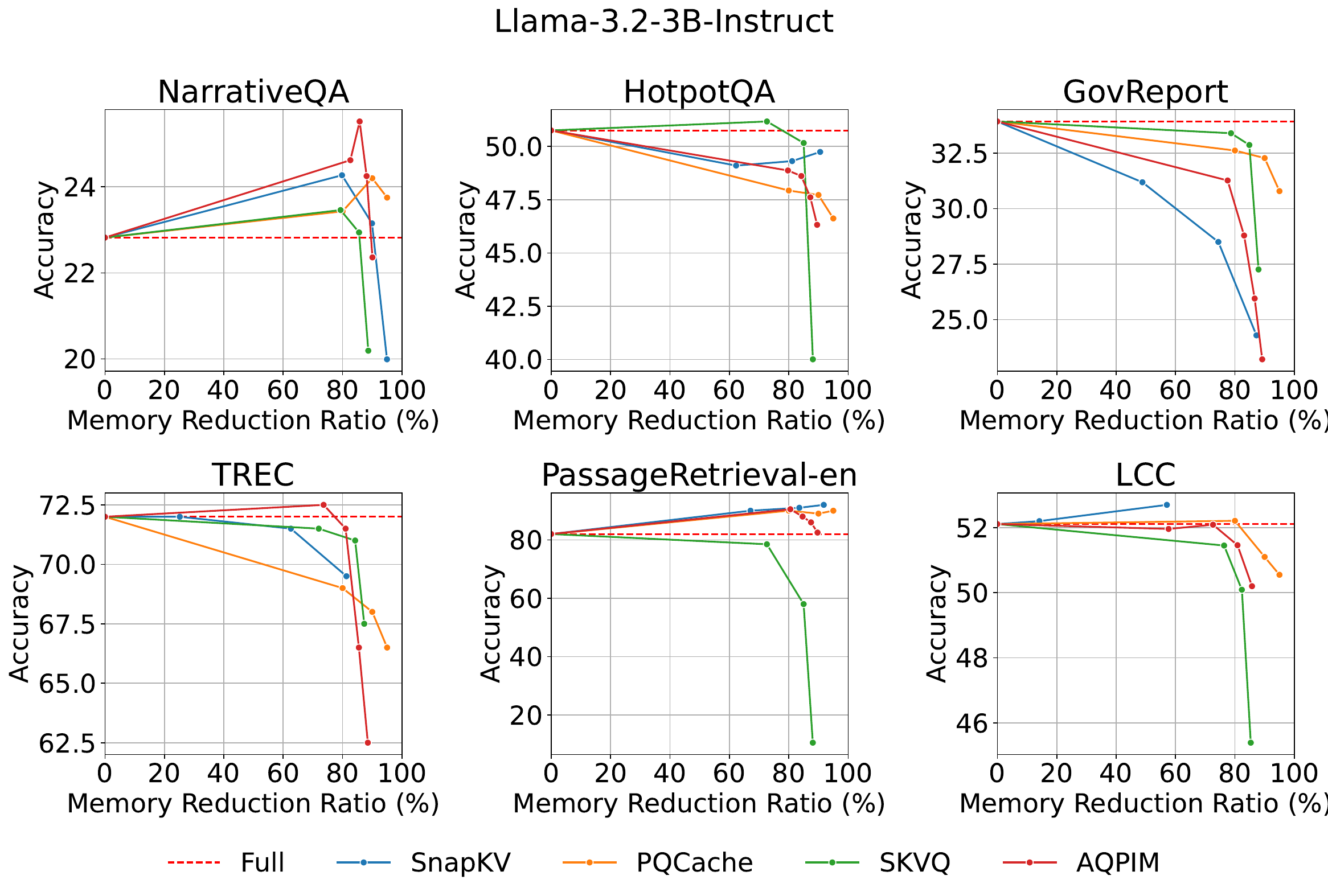}
    \end{subfigure}
    \caption{Memory reduction ratio vs. accuracy.}
    \label{fig:tradeoff}
    \vspace{-3mm}
\end{figure*}

\subsection{Experiments on LongBench}
We compare the tradeoff between memory reduction ratio and accuracy of \AQPIM against SnapKV, PQCache, and SKVQ. As shown in~\cref{fig:tradeoff}, \AQPIM achieves a comparable tradeoff across all tasks and on both models.
Compared to SnapKV and SKVQ, \AQPIM tends to exhibit a better trade-off, thereby demonstrating that PQ, a clustering-based method, delivers higher quantization quality. While PQCache maintains high accuracy even under aggressive compression ratios by storing the full KV cache in CPU memory and thereby mitigating information loss, \AQPIM achieves comparable accuracy up to approximately 80\% compression while operating entirely within PIM memory.

\subsection{Ablation Study}
To evaluate the effectiveness of the importance-weighted k-means clustering and vector splitting optimization, we conduct ablation studies comparing four configurations: standard PQ, \AQPIM without weighting, \AQPIM without pre-sorting, and \AQPIM.
\cref{tb:ablation_study} presents the result for high compression scenarios (128 centroids). We observe that applying both weighting and pre-sorting significantly contributes to accuracy, particularly under aggressive compression situations. The overhead introduced by these techniques is minimal: importance-weighted k-means leverages the attention scores generated during attention computation, and the channel permutation is generated offline using a calibration dataset and integrated into the projection weights during inference.
\begin{table}[tb]
    \centering
    \caption{
    Effect of introduced PQ optimizations. Both importance-weighted clustering and channel pre-sorting contribute to accuracy improvement particularly under aggressive compression situations.}  
    
    \label{tb:ablation_study}
    \begin{tabular}{c|cccc}
        \bhline{1pt}
        Configuration & Standard PQ & w/o weighting & w/o pre-sort & \AQPIM \\
        \hline
        NarrativeQA & 18.35 & 20.51 & 21.42 & 23.09 \\
        HotpotQA & 37.06 & 34.39 & 36.65 & 38.08 \\
        GovReport & 24.45 & 25.63 & 23.55 & 25.49 \\
        TREC & 70.65 & 70.15 & 69.65 & 70.50 \\
        PRetrieval & 61.26 & 55.49 & 86.47 & 88.17 \\
        LCC & 53.94 & 53.34 & 54.83 & 54.66 \\
        \hline
        Average & 44.29	& 43.25	& 48.76	& \textbf{50.00} \\
        \bhline{1pt}
    \end{tabular}
\end{table}

\begin{figure*}
    \centering
    \includegraphics[width=\linewidth]{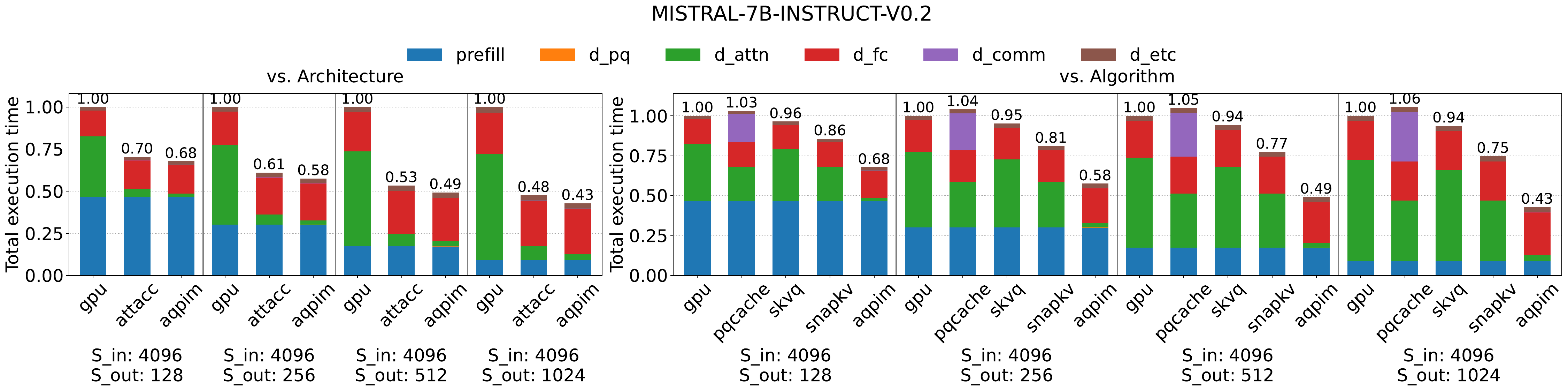}
    \caption{Normalized total execution time comparing different architectures (left) and algorithms (right).}
    \label{fig:lout}
    \vspace{-1mm}
\end{figure*}

\begin{figure*}
    \centering
    \includegraphics[width=\linewidth]{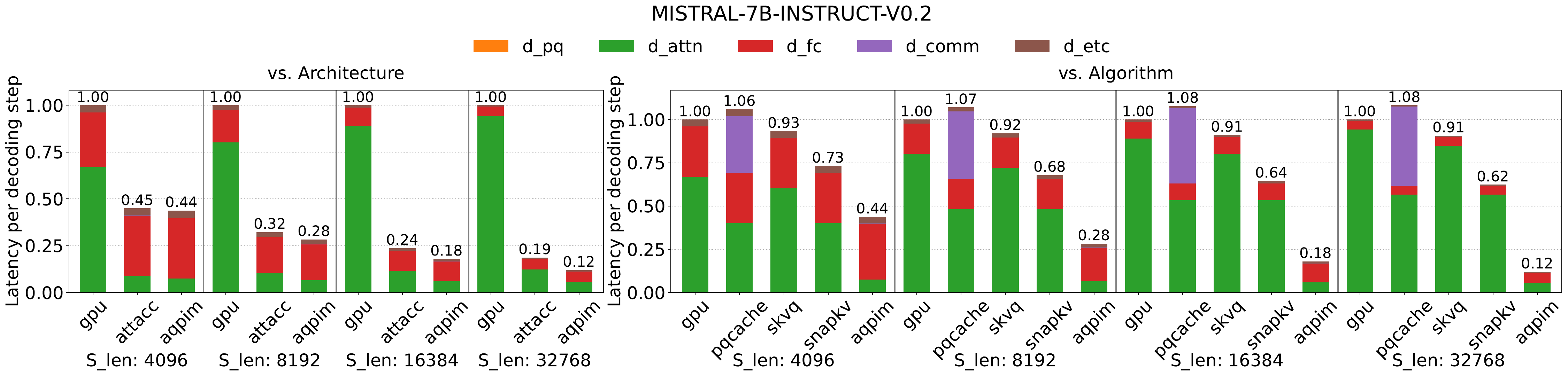}
    \caption{Normalized decoding time comparing different architectures (left) and algorithms (right).}
    \label{fig:lin}
    \vspace{-1mm}
\end{figure*}

\subsection{System Simulation}
\label{ssec:system_simulation}
\textbf{Hardware configuration}:
We simulate the performance of the \AQPIM by comparing a conventional system with GPU+HBMs, GPU+HBM-PIMs (AttAcc!~\cite{park2024attacc!}), GPU+\AQPIM. All simulated configurations utilize NVIDIA's H100 GPUs~\cite{nvidia2024h100}.
The baseline conventional GPU+HBMs consists of 1 H100 GPU core and 5 16GB HBM modules. In the PIM-enabled systems (AttAcc! and \AQPIM) replace 4 16GB HBMs with 4 16GB HBM-PIMs. These HBM-PIMs are allocated for KV cache storage, while the remaining HBMs are used for other data, such as model parameters. In addition, we used Intel's Xeon Platinum 8480+ Processor~\cite{intel2023intel} in our experiments with CPU. 


\textbf{Baselines}:
We compare \AQPIM against other architectures and KV cache compression methods.
The architectures include GPU+HBMs and AttAcc!, while
the KV cache compression methods include PQCache, SKVQ, and SnapKV. For \AQPIM, we set the number of subvectors to 32 and the number of centroids to 512. For the other compression methods, we set the memory reduction ratio to 80\%.

Note a key difference in assumptions: AttAcc! does not inherently address scenarios where large KV caches overflow the HBM-PIM memory.
In contrast, \AQPIM operates under the assumption that while the model itself can fit within memory, the full KV cache may not, thus necessitating compression.
Importantly, our approach can also scale out to provide a larger memory capacity to accommodate larger models. 
Even when the dimensionality increases, \AQPIM can scale by increasing the number of subvector sets to keep the per-codebook vector space constant. This preserves the expressive power of subvectors and codebooks while enabling scaling with the abundant bank resources. 

\textbf{Simulator:}
To evaluate these baselines, we constructed a customized GPU-PIM simulator for LLM inference, built upon the AttAcc! simulator~\cite{scale-snu-attacc_simlator}. This simulator itself is a modified version of Ramulator~\cite{yoongu2016ramulator, haocong2024ramulator2.0, safarirg2023ramulator2.0}, adapted to simulate LLM inference. The simulator takes three inputs: system configuration, model details, and input configurations, and outputs both execution time and energy consumption.

\RBtext{
\textbf{Energy and Area:} 
\AQPIM repurposes most of the\RB{B3}\label{RB:B3} existing HBM peripherals (e.g., 1K row buffer and column decoder) and the PE designs implemented in AttAcc!~\cite{park2024attacc!}. The added logic for intra-row indirection, which is synthesized with the same PDK~\cite{ASAP7} and scaled with the same DRAM die density ratio as AttAcc!, is 0.0565$\mathrm{mm}^2$ per HBM. This is only 0.43\% of BankPE area 
of HBM3 implementing AttAcc!. BufferPE in AttAcc! already has all the required components in~\Cref{tb:operation_breakdown} in their accumulators and softmax units.  

Timing and energy consumption used in our simulation are based on the synthesis results reported on AttAcc!. Since the column decoder input is latched for pipelining, $t_\text{CCDL}$ (delay between RDs to the same BG) will not be affected by our modifications. DRAM traffic and contentions are entirely managed by DRAM controllers and modeled in our simulator. 

}

While our tested model and algorithm employ the de facto standard bfloat16, the architecture simulation is conducted using FP16 to make it directly comparable to prior work~\cite{park2024attacc!}. Comparing FP16 and bfloat16 MAC implementations on HBM-PIM, the bfloat16 unit has identical latency but provides 13\% smaller area and 14\% better energy efficiency per operation~\cite{lee2021hardware}. While FP16 can still gain considerable benefits from our approaches, bfloat16 would also be a compelling option for future HBM-PIM designs.


\begin{figure}[tb]
    \centering
    \includegraphics[width=0.99\linewidth]{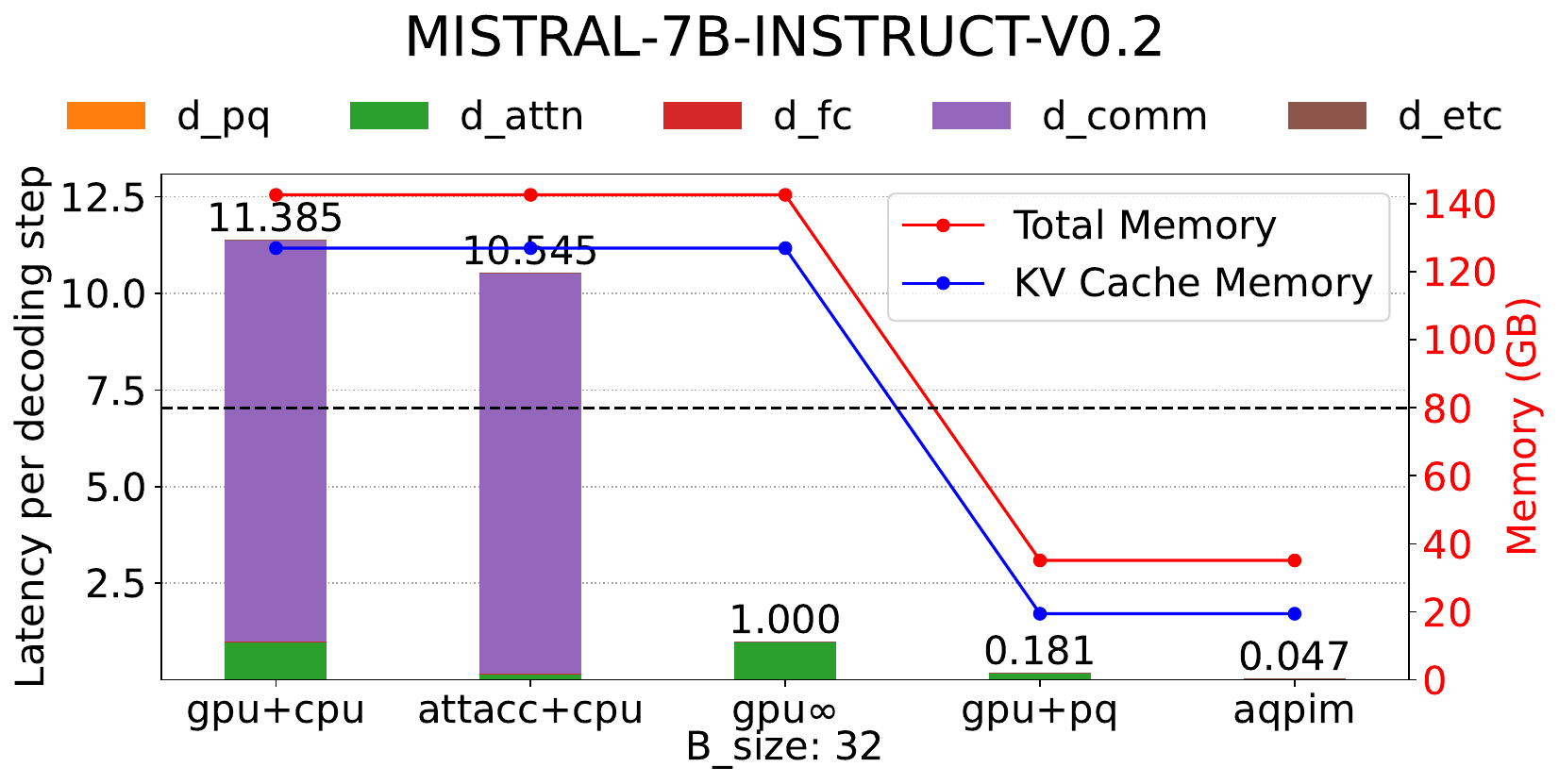}
    \caption{\RBtext{Decomposition analysis of decoding speedups. \texttt{gpu$\infty$} is a baseline that assumes infinite GPU memory capacity. Both \texttt{gpu+pq} and \texttt{aqpim} utilize PQ compression.}
    }
    \label{fig:overflow}
    \vspace{-3mm}
\end{figure}

\begin{figure*}
    \centering
    \includegraphics[width=\linewidth]{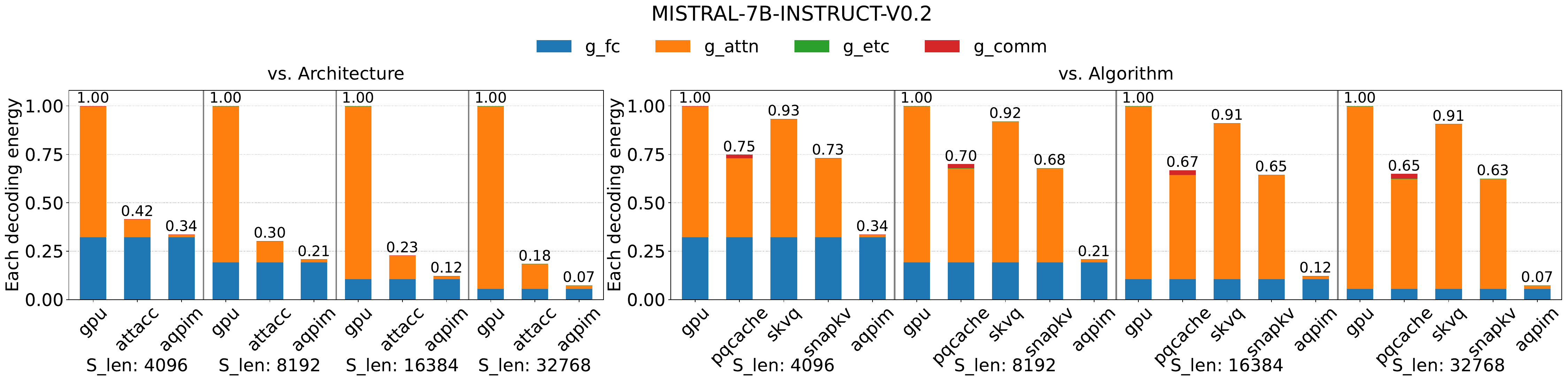}
    \caption{Normalized energy for decoding comparing different architectures (left) and algorithms (right).}
    \label{fig:energy}
    \vspace{-2mm}
\end{figure*}

\textbf{Scenarios \& Models:}
Our experiments evaluate a range of scenarios by varying input lengths, output lengths, and batch sizes. Unless otherwise specified, we use a batch size of 16. 
The evaluated model is Mistral-7B-Instruct-v0.2~\cite{Jiang2023mistral7b}.

\subsection{Performance}
\label{ssec:performance}

We first compare the total execution time of the different approaches. \cref{fig:lout} presents the normalized total execution time for a 4K input, varying the output length. The left graph compares the performance of GPU, AttAcc!, and \AQPIM.
\AQPIM significantly reduces the matrix multiplication (matmul) execution time during the decoding phase. This reduction ratio increases with longer output lengths, achieving up to a 2.33$\times$ faster overall execution time.

The right graph compares PQCache, SKVQ, SnapKV, and \AQPIM.
All compression methods, except for \AQPIM, were executed on GPU, since their designs are not primarily intended for PIM. As illustrated, PQCache experiences performance limitations due to communication overhead with CPU memory. SKVQ achieves some speedup over the GPU baseline due to the reduced data transfer, but its acceleration is limited since current GPUs do not efficiently handle quantized low bit values, often requiring upcasting to larger bit precision. SnapKV demonstrates better acceleration than other methods, but its performance gain remains considerably smaller than that of \AQPIM. This is attributed to \AQPIM's ability to leverage both architectural acceleration from PIM and algorithmic optimizations from PQ-based attention.

Moreover, as the output length increases, the decoding phase becomes dominant in the total execution time. Therefore, our subsequent analysis focuses solely on the decoding process.


We secondly compare the execution time of each decoding step, varying the input length. \cref{fig:lin} presents the normalized execution time per decoding step.
The left graph compares different architectures, and the right graph compares algorithms.
\AQPIM greatly reduces matmul execution time compared to both architectures and algorithms, achieving up to an 8.33$\times$ speedup, and this effect becomes more notable as the input length increases. This is because \AQPIM maintains a fixed number of centroids at 512, which fit within a DRAM row, thus ensuring a constant matmul cost. Although the retrieval cost increases with sequence length, this cost remains negligible due to our efficient intra-row indirection mechanism, as discussed in \Cref{ssec:system_architecture}.

\RBtext{
We\RB{A1}\label{RB:A1} further analyze the speedup breakdown of \AQPIM. In this analysis, we will also assume an imaginary GPU with infinite memory capacity (\texttt{gpu$\infty$}) to separate the contribution from GPU-CPU communication reduction of \AQPIM. \cref{fig:overflow} presents four evaluation scenarios: \texttt{gpu+cpu} (GPU offloads KV cache to CPU when it overflows), \texttt{gpu$\infty$} (infinite GPU memory, no offloading penalty), \texttt{gpu+pq} (PQ compressed KV with GPU), and \texttt{\AQPIM}. 
Note that \texttt{gpu+pq} does not account for PQ compression overhead incurred during prefilling, thus providing an idealistic result. 

The elimination \RB{A1} of the offloading penalty yields a performance gain of 11.39$\times$. This roughly aligns with the gap between GPU's memory bandwidth (3.35TB/s) and PCIe bandwidth (256GB/s). The reduced memory footprint from PQ compression contributes to an additional 5.52$\times$ speedup. This is consistent with the high data compressibility of PQ (6.53$\times$ reduction of KV capacity). Finally, \AQPIM's dedicated architectural optimization yields another 3.85$\times$ speedup. 
While our PIM baseline, AttAcc!, offers 7.2$\times$ higher aggregated internal memory bandwidth compared to GPU (3.35TB/s), the un-accelerated operations (e.g. FFN) dominate the decoding latency of \AQPIM. However, solely comparing the attention phase observes 10.33$\times$ speedup, exceeding the bandwidth gap. This happens because \AQPIM can produce more data than a row with a single ACT through data reuse in row buffers during indirection. 

When compared to an imaginary AttAcc! (not shown in the figure) that has infinite capacity but the same PE counts, \AQPIM achieves 3.4$\times$ speedup. While the gain can be explained by the reduced KV size, it is smaller than \texttt{gpu+pq}'s 5.52$\times$ speedup against \texttt{gpu$\infty$} because AttAcc's PIM diminishes the share of attention in the total decoding time.
}

\RBtext{
Lastly, \RB{C2}\label{RB:C2} we evaluate the benefits of intra-row indirection. An alternative approach is to perform gather operations on the BufferPE, transferring the entire row data. The results are summarized in \cref{tb:intra-row_indirection}. While intra-row indirection at BankPEs significantly reduces off-bank data transfer, performing it on the BufferPE incurs a costly data transfer and resource contention. While the gap is narrower for Keys as they are anyway transferred to BufferPEs for softmax operations, value matrix experiences significant overhead as it necessitates unnecessary roundtrip to the BufferPE between BankPE operations.
}

\begin{table}[t]
    \centering
    \caption{
    \RBtext{Indirection cycles (Sequenth length = 4K).}
    }
    \vspace{-2mm}
    \label{tb:intra-row_indirection}
    {\footnotesize
    \begin{tabular}{c|c|c}
        \bhline{1pt}
          & On BankPE & On BufferPE \\
         \hline
        Key (including transfer to Softmax) & 33089 & 37185 \\
        Value & 7373 & 181875 \\
        \bhline{1pt}
    \end{tabular}
    }
\end{table}

\begin{figure}[tb]
    \centering
    \includegraphics[width=\linewidth]{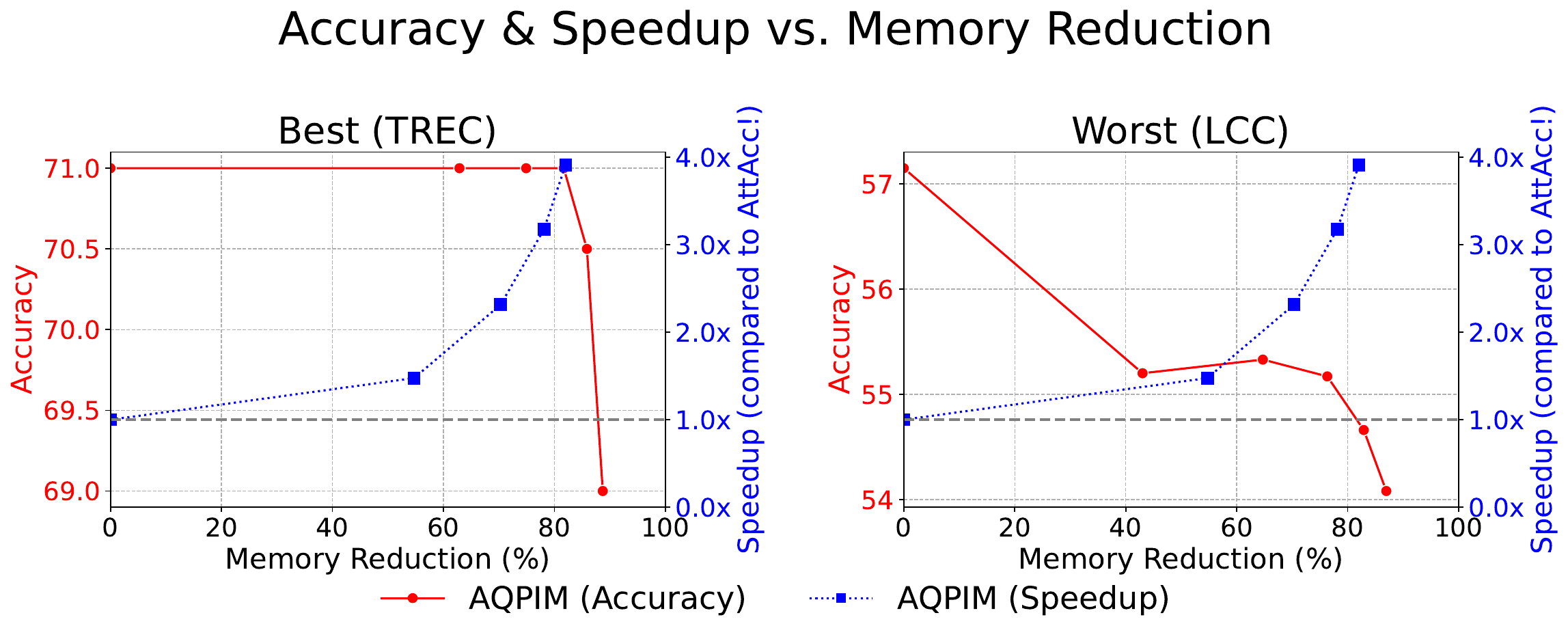}
    \caption{
    \RBtext{Accuracy \& Speedup vs. Memory Reduction.}
    }
    \vspace{-3mm}
    \label{fig:acc&speed_vs_reduction}
    \vspace{-3mm}
\end{figure}

\RBtext{
\label{RB:D3}
\subsection{Accuracy \& Speedup vs. Memory Reduction}
\cref{fig:acc&speed_vs_reduction} analyzes\RB{D3} the the trade-offs among accuracy, speedup, and memory reduction focusing on the attention kernel. Due to limited space, we show those with the best and worst tradeoffs. In the best-case scenario (left), \AQPIM maintains high accuracy even with extreme compression, achieving a significant speedup. Even in the worst case scenario, \AQPIM achieve a comparable accuracy while offering a substantial speedup. Note that compression rate can be flexibly adjusted without hardware modification. 
}

\subsection{Energy Efficiency}
\label{ssec:energy_efficiency}

We estimated the energy consumption across different approaches. \cref{fig:energy} presents the normalized energy consumption per decoding step, varying the input length. 
Regarding architectural comparisons (left), \AQPIM significantly reduces the energy consumed in attention calculation. Compared to the GPU baseline, \AQPIM achieves up to a 14.29$\times$ improvement in energy efficiency.
In terms of algorithmic comparisons (right), \AQPIM surpasses other methods in energy efficiency for 
``attention'' components.
The energy for "attention" is decreased because \AQPIM utilizes a hardware-aware PQ-based attention mechanism, incorporating fixed-size matmul and row buffer retrieval mechanism.

%% file: sections/6_conclusion.tex
\section{Conclusion}
\label{sec:conclusion}

In this work, we observe that activations exhibit both locality and similarity, derived from contextual information. Based on the trade-off between compressibility and required memory bandwidth, we propose a clustering-based approach to better exploit the underutilized capabilities of PIM. PQ, a clustering-based method, is well-suited to preserve locality. We further enhance PQ by introducing importance-aware and data-driven optimizations to maintain high accuracy while drastically reducing memory footprint. Additionally, we propose a fast attention computation mechanism that directly operates on the compressed format, enabling localized memory access patterns favorable for PIM. 